\documentclass[11pt, a4paper, logo, copyright]{googledeepmind}

\usepackage[authoryear, sort&compress, round]{natbib}

\usepackage{cleveref}
\usepackage{url}
\usepackage{wrapfig,booktabs}
\usepackage{xcolor}         % colors
\usepackage{tikz}
\usetikzlibrary{patterns}
\usepackage{colortbl}
\usepackage{arydshln}
\usepackage{adjustbox}
\usepackage{multirow}
\usepackage{enumerate}
\usepackage{enumitem}
\usepackage{subfigure}
\usepackage{gensymb}
\usepackage{xspace} % for algorithm names
\usepackage{makecell}  % Include this in the preamble
\usepackage{minitoc}             % For creating the appendix TOC
\usepackage{makeidx}
\usepackage{booktabs}
\usepackage{graphicx}

\usepackage{bbm}
\usepackage{enumerate}
\usepackage{algorithm}
\usepackage{algorithmic}
\usepackage{colortbl}
\usepackage{psfrag}
\usepackage{adjustbox}
\usepackage{wrapfig}
\usepackage{xspace}
\usepackage{amssymb}
\usepackage{multirow}
\usepackage{afterpage}
\usepackage{float}
\usepackage[toc,page,header]{appendix}
\usepackage{minitoc}
\usepackage{placeins} % For FloatBarrier
\usepackage{titletoc} % Required for partial ToCs
\colorlet{darkgreen}{green!65!black}
\colorlet{darkblue}{blue!75!black}
\colorlet{darkred}{red!80!black}
\definecolor{lightblue}{HTML}{0071bc}
\definecolor{lightgreen}{HTML}{39b54a}
\definecolor{manyshot}{HTML}{6969ff}
\definecolor{medshot}{HTML}{f7c600}
\definecolor{fewshot}{HTML}{ff6969}
\definecolor{mypurple}{HTML}{412F8A}
\definecolor{myorange}{HTML}{fc8e62}

\definecolor{deemph}{gray}{0.55}

\definecolor{textgreen}{RGB}{57, 172, 57}
\definecolor{textred}{RGB}{200, 10, 10}
\definecolor{textgray}{RGB}{100, 100, 100}
\definecolor{visiongold}{RGB}{230, 184, 0}
\definecolor{speechpurple}{RGB}{204, 0, 255}
\definecolor{dataprep}{RGB}{38, 189, 128}
\definecolor{modeltraining}{RGB}{38, 189, 128}
\definecolor{backgroundcol}{RGB}{232, 230, 230}
\definecolor{gold}{rgb}{225, 215, 200} % Gold color
\definecolor{navyblue}{RGB}{40, 66, 200} % Navy blue color
\definecolor{orange}{RGB}{255,127,80} % Navy blue color
\definecolor{pink}{RGB}{219,112,147} % Navy blue color

\definecolor{baselinecolor}{gray}{.95}

\usepackage{pifont}% http://ctan.org/pkg/pifont

\usepackage{makecell, tabularx}
\newcolumntype{L}{>{\RaggedRight}X}
\usepackage{siunitx}

\title{Passive Heart Rate Monitoring During Smartphone Use in Everyday Life}

\correspondingauthor{$mingzher$@google.com}

\author[$\circ$,1]{Shun Liao}
\author[$\circ$,1]{Paolo Di Achille}
\author[$\circ$,1]{Jiang Wu}
\author[$\circ$,1]{Silviu Borac}
\author[$\circ$,1]{Jonathan Wang}
\author[$\circ$,1]{Xin Liu}
\author[$\circ$,1]{Eric Teasley}

\author[1]{Lawrence Cai}
\author[1]{Yuzhe Yang}
\author[1]{Yun Liu}
\author[1]{Daniel McDuff}
\author[1]{Hao-Wei Su}
\author[1]{Brent Winslow}
\author[1]{Anupam Pathak}
\author[1]{Shwetak Patel}
\author[1]{James A. Taylor}

\author[$\ddagger$,1]{Jameson K. Rogers}
\author[$\dagger$,$\ddagger$,1]{Ming-Zher Poh}

\affil[$\circ$]{Equal contributions}
\affil[$\ddagger$]{Equal leadership}
\affil[$\dagger$]{Corresponding Author}
\affil[1]{Google Research}

\begin{abstract}
Resting heart rate (RHR) is an important biomarker of cardiovascular health and
mortality, but tracking it longitudinally generally requires a wearable device,
limiting its availability. We present PHRM, a deep learning system for passive
heart rate (HR) and RHR measurements during everyday smartphone use, using facial
video-based photoplethysmography. Our system was developed using 225,773 videos
from 495 participants and validated on 185,970 videos from 205 participants in
laboratory and free-living conditions, representing the largest validation study
of its kind. Compared to reference electrocardiogram, PHRM achieved a mean
absolute percentage error (MAPE) $<10$\% for HR measurements across three skin
tone groups of light, medium and dark pigmentation; MAPE for each skin tone
group was non-inferior versus the others. Daily RHR measured by PHRM had a mean absolute
error $<5$ bpm compared to a wearable HR tracker, and was associated with known
risk factors. These results highlight the potential of smartphones to enable passive
and equitable heart health monitoring.
\end{abstract}

\begin{document}

\maketitle

\newenvironment{Itemize}{
    \begin{itemize}[leftmargin=*]
    \setlength{\itemsep}{0pt}
    \setlength{\topsep}{0pt}
    \setlength{\partopsep}{0pt}
    \setlength{\parskip}{0pt}}
{\end{itemize}}
\setlength{\leftmargini}{9pt}

\section{Introduction}

Heart rate (HR) is an important and dynamic vital sign that varies based on numerous influences including physical and mental activity, sleep stages, and environmental factors \citep{Ceconi2011-iw}. In addition, resting heart rate (RHR) is well-recognized as a biomarker of cardiovascular health and a prognostic for overall mortality \citep{Kannel1987-ul, Raisi-Estabragh2020-sf, Alhalabi2017-ws}. Studies of longitudinal changes in RHR over long periods of time have reported that temporal increases in RHR were associated with higher mortality and major adverse cardiovascular events \citep{Nauman2011-wt, Seviiri2018-oj, Vazir2018-sm}. The traditional method of measuring RHR relies on measurements taken after a period of rest, typically in a supine or sitting position. While relatively simple to perform, this limits the practicality of obtaining daily RHR measurements over time to evaluate longitudinal trajectories. As such, until the proliferation of consumer wearable devices equipped with HR sensors, the quantification of RHR was largely confined to clinical and research settings, with limited observation of RHR during individuals’ daily lives. Considering the sensitivity of HR to various factors, the cardiovascular system might be better assessed through multiple measurements during daily life activities rather than brief measurements during standardized rest in an artificial and often stressful clinical environment \citep{Johansen2013-qd, Hansen2007-pb, Dunn2021-wy}. Indeed, average HR throughout the day has been shown to be a strong independent predictor, even more so than RHR, for all-cause mortality \citep{Korshoj2015-rc}. In addition, continuously measuring ambulatory HR throughout a day has demonstrated improved reproducibility compared to clinic-based RHR \citep{Dunn2021-wy, Palatini2000-hh}.

With the availability of consumer wearable devices integrated with photoplethysmogram (PPG) sensors, daily RHR derived from combining multiple HR measurements throughout the day can now be automatically tracked \citep{Russell2019-wh}. Daily RHR tracking might provide a measure of an individual’s overall cardiovascular status and capture changes over weeks in cardiovascular fitness, or over days due to illness or other significant physiological triggers \citep{Quer2020-xk, Alexander2022-gt, Mishra2020-sx}. However, the adoption and consistent utilization of consumer wearable devices remains limited, and there is lower adoption by those who are most likely to benefit from these technologies \citep{Dhingra2023-ui}. Smartphones provide an attractive alternative to develop daily RHR monitoring capabilities as the vast majority (90\%) of US adults already have a smartphone and global smartphone penetration is estimated at 69\% \citep{Gelles-Watnick2024-bx}. Moreover, Americans check their phones an average of 144 times per day and the majority use their phones within the first 10 minutes of waking up \citep{Kerai2023-kz}. In our work, we leverage the smartphone to provide a platform for opportunistically measuring HR across the day during normal phone use and aggregate the HR measurements across the day to derive RHR.

\begin{figure*}[t!]
	\centering
	\includegraphics[width=0.95\textwidth, keepaspectratio]{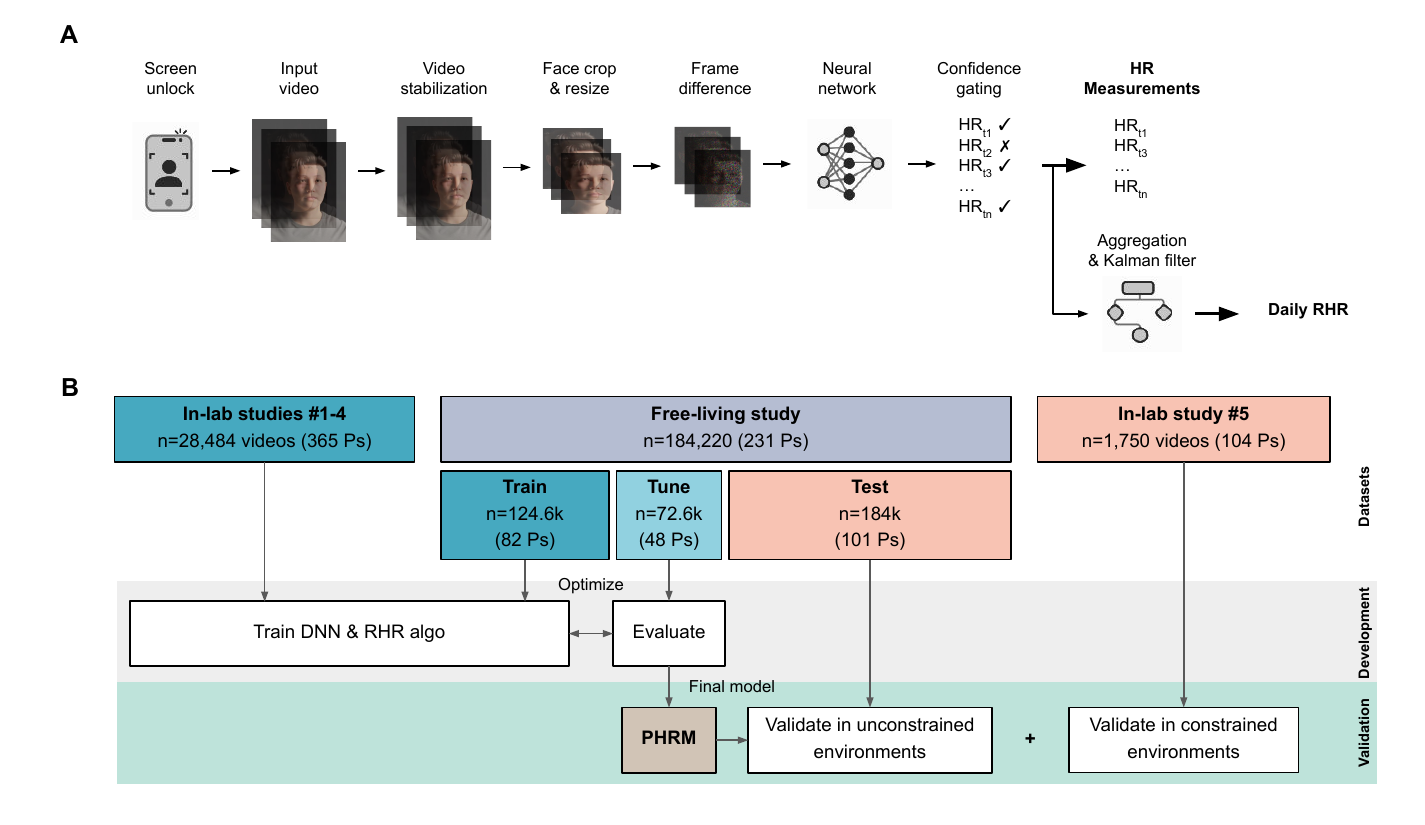}
	\caption{\textbf{System overview, development, and validation of the deeplearning system for passive heart rate (HR) and daily resting HR (RHR) measurements (PHRM) during smartphone use.} (A) In our research study with consented participants, upon a screen unlock event, the PHRM passively captures, processes, and analyzes 8-sec facial video using a deep neural network (DNN) to estimate HR and associated prediction confidence to determine if the measurement	is valid. To compute daily RHR, the PHRM aggregates valid HR measurements from intermittent 8-sec video clips throughout a single day and applies a Kalman filter to improve estimates. (B) Workflow diagram of the different studies used to develop and validate the PHRM system. We used data from five independent,	prospective laboratory studies and a prospective free-living study to develop and validate the PHRM.}
	\label{fig:figure_1}
\end{figure*}

The principles of measuring the blood volume pulse from a distance via video-based, remote photoplethysmography (rPPG) are well established \citep{Sun2016-jm, wang2016algorithmic}. rPPG has been demonstrated to be capable of measuring HR \citep{poh2017validation, Yan2017-st, Qiao2022-fj, Liu2020-tg} and screening for irregular heart rhythms such as atrial fibrillation \citep{Yan2018-vi} via analysis of video images of the human face captured using using front-facing smartphone cameras. However, most rPPG studies to date comprised small sample sizes or were conducted in controlled environments, limiting the generalizability to real-world conditions. Furthermore, the accuracy of current rPPG methods drops significantly for the darkest skin tones due to a larger concentration of melanin which absorbs more light \citep{Nowara2020-fy}. The growing concern that accuracy of optical devices using PPG, particularly pulse oximeters, may vary by skin pigmentation has led to increased scrutiny from health governing bodies including the U.S. Food \& Drug Administration (FDA) \citep{Center_for_Devices_and_Radiological_Health2023-hy} and NHS \citep{Department_of_Health_and_Social_Care2024-ui}. This has prompted recommendations for clinical validation studies to be conducted on a large number of participants and a diverse range of skin pigmentation. Thus far, there is a lack of rPPG studies that would meet the diversity requirements proposed by the FDA in terms of distribution and range of skin pigmentation \citep{Center_for_Devices_and_Radiological_Health2023-hy}.

In this work, we present, for the first time, a smartphone-based, deep learning system that enables passive measurements of HR and daily RHR in the background during normal phone use (collectively referred to as passive heart rate monitoring; PHRM). Compared to previous work, our system provides several advances. First, we validated its performance in a prospective study on the largest and most diverse set of videos ($>185$k) to date, collected in laboratory conditions as well as in free-living, real-world conditions using participants’ personal phones. Second, we demonstrate how our system provides equitable and accurate HR and daily RHR readings across all skin pigmentation groups. Finally, we demonstrate that PHRM-derived daily RHR is associated with well-established cardiovascular health metrics and risk factors.
\section{Results}

\begin{table}[!t]
	\centering
	\caption{\textbf{Baseline characteristics of participants across studies.}}
	\resizebox{\textwidth}{!}{\begin{tabular}{lllllll}
	\toprule                                         &                                                                                   & \multicolumn{2}{l}{\textbf{Train}} & \textbf{Tune} & \multicolumn{2}{l}{\textbf{Test}} \\
	Conditions                                       &                                                                                   & Laboratory                         & Free-living   & Free-living                      & Laboratory  & Free-living \\
	\midrule \multicolumn{2}{l}{No. videos}          & 28,484                                                                            & 124,683                            & 72,606        & 1,750                            & 184,220      \\
	\multicolumn{2}{l}{No. participants}             & 365                                                                               & 82                                 & 48            & 104                              & 101          \\
	\multicolumn{2}{l}{Heart rate, BPM (mean ± STD)} & 92.8 ± 32.0                                                                       & 66.9 ± 34.5                        & 69.3 ± 29.1   & 74.6 ± 13.4                      & 70.2 ± 29.8  \\
	\multicolumn{2}{l}{Age (mean ± STD)}             & 40.5 ± 14.7                                                                       & 37.9 ± 9.8                         & 37.4 ± 11.8   & 51.3 ± 14.8                      & 38.1 ± 11.4  \\
	\multirow{3}{*}{Age group}                       & \textless{}40 years                                                               & 196 (54.1\%)                       & 52 (61.9\%)   & 32 (66.7\%)                      & 24 (23.1\%) & 62 (57.4\%) \\
	                                                 & 40-59 years                                                                       & 113 (31.2\%)                       & 30 (35.7\%)   & 12 (25.0\%)                      & 43 (41.3\%) & 42 (38.9\%) \\
	                                                 & \textgreater{}59 years                                                            & 53 (14.6\%)                        & 2 (2.4\%)     & 4 (8.3\%)                        & 37 (35.6\%) & 4 (3.7\%)   \\
	\multicolumn{2}{l}{No. female (\%)}              & 199 (55.0\%)                                                                      & 45 (53.6\%)                        & 25 (52.1\%)   & 71 (68.3\%)                      & 58 (53.7\%)  \\
	\multirow{3}{*}{Skin pigmentation group: n (\%)} & \begin{tabular}[c]{@{}l@{}}Group 1\\ (Fitzpatrick I-III; \\ MST 1-4)\end{tabular} & 165 (45.6\%)                       & 30 (35.7\%)   & 18 (37.5\%)                      & 44 (42.3\%) & 40 (37.0\%) \\
	                                                 & \begin{tabular}[c]{@{}l@{}}Group 2\\ (Fitzpatrick IV-V; \\ MST 5-7)\end{tabular}  & 116 (32.0\%)                       & 22 (26.2\%)   & 12 (25.0\%)                      & 25 (24.0\%) & 29 (26.9\%) \\
	                                                 & \begin{tabular}[c]{@{}l@{}}Group 3\\ (Fitzpatrick VI; \\ MST 8-10)\end{tabular}   & 81 (22.4\%)                        & 32 (38.1\%)   & 18 (37.5\%)                      & 35 (33.7\%) & 39 (36.1\%) \\
	\bottomrule
\end{tabular}} \label{tab:baseline}
	\parbox{\textwidth}{\raggedright \footnotesize MST represents Monk skin tone.}
\end{table}

\paragraph{Overview of System}
We designed and developed the PHRM with two major components Fig.~\ref{fig:figure_1}.
First, we constructed an end-to-end HR estimation module that takes as input a
short (8-second) video clip of the user’s face, performs video stabilization,
pre-processing (by face cropping, resizing, interpolating and computing frame differences) and predicts
HR along with a measure of confidence using an ensemble \citep{ganaie2022ensemble}
of computationally-efficient temporal shift convolutional neural networks (TS-CNNs)
\citep{Liu2020-tg}. Next, we designed an algorithm to derive daily RHR by aggregating
the HR predictions throughout the day using the confidence of predictions and a Kalman
filter. The PHRM was designed to run passively in the background and
automatically initiate video capture via the front-facing camera upon a screen
unlock event.

\subsection{Study Populations}
To develop and validate the PHRM, we conducted a series of studies to acquire datasets
comprising face videos and HR ground truth (Table~\ref{tab:baseline}). In all
our studies, we recruited for diversity across age, sex and skin tone groups. We
used the electrocardiogram (ECG) as the reference HR ground truth for both the laboratory-based (in-lab)
and free-living validation studies. In total, we collected 225,773 videos from 495
participants for PHRM development, and 185,970 videos from 205 participants for
PHRM validation.

First, we obtained data to train and tune the PHRM from four separate studies performed
in controlled laboratory settings (n=28,484 videos from 365 participants). This
data comprises a variety of lighting conditions and physiologic states including
at rest, during various exercises, and post-exercise (details in Table S~\ref{tab:hr3_result}).
To provide an external test set for model validation, we conducted a fifth, prospective
laboratory study that enrolled 104 participants (n=1,750 videos) and captured videos
under five different lighting conditions and both at-rest and post-exercise physiologic
states. The mean age in this external test set was 51.3 ± 14.8 years; 71 (68.3\%)
participants were female. We divided participants into three groups of skin pigmentation
(Fitzpatrick I-III, Fitzpatrick IV-V, Fitzpatrick VI) based on converting their objective
individual topology angle (°ITA) - as measured by a spectrocolorimeter at the cheeks
and forehead - Fitzpatrick skin tones \citep{del2013variations}. We specified these
skin tone groups to intentionally overrepresent participants of the darkest skin
tones and ensure development of models that perform accurately for this group, a
decision that aligned with the three skin pigmentation cohorts subsequently
proposed by the FDA \citep{Center_for_Devices_and_Radiological_Health2023-hy}.
ITA values ranged from -73.48° to 88.81° with 44 (42.3\%), 25 (24.0\%), and 35 (33.7\%)
participants in skin pigmentation group 1 (lightest), 2 (medium) and 3 (darkest),
respectively.

Next, we conducted a prospective free-living study designed to passively record face
videos during normal phone use over an 8-day period. The detailed video
recording protocol is provided in Appendix~\ref{app:freeliving_protocal}. We applied
stratified sampling based on age, sex, body mass index (BMI), and Monk skin tone (MST) to split the free-living data at the participant level: data from 50\% of the
participants (n=197,289 videos from 130 participants) were set aside for model
development (30\% for training and 20\% for tuning), and data from the remaining
50\% of participants (n=184,220 videos from 101 participants) were set aside as
the test split for validation. We switched to using the MST in the prospective free-living study since it was designed to be more inclusive of the spectrum of skin tones we see in our society (the laboratory studies were conducted before the introduction of MST and used Fitzpatrick skin tone, the de-facto industry standard at that time). The 101 participants in the test split of the free-living
study excludes six individuals who did not meet the minimum adherence criteria,
i.e. at least 3 days with more than 40 video clips per day (Fig. S~\ref{fig:flow_chart}).
The mean age was 38.1 ± 11.4 years; 58 (53.7\%) participants were female.
Following FDA’s proposal, the entire range of skin pigmentation based on the self-reported
MST was represented with at least one participant for each MST value
of 1-10. We divided participants into three MST cohorts, yielding 40 (37.0\%),
29 (26.9\%), and 39 (36.1\%) participants in MST 1-4, MST 5-7 and MST 8-10 cohorts,
respectively. This distribution also fulfilled the FDA recommendation to have $\ge
40$\% of each sex and $\ge 25$\% of participants in each of the 3 MST cohorts.

\begin{figure*}[!t]
	\centering
	\includegraphics[width=1 \textwidth]{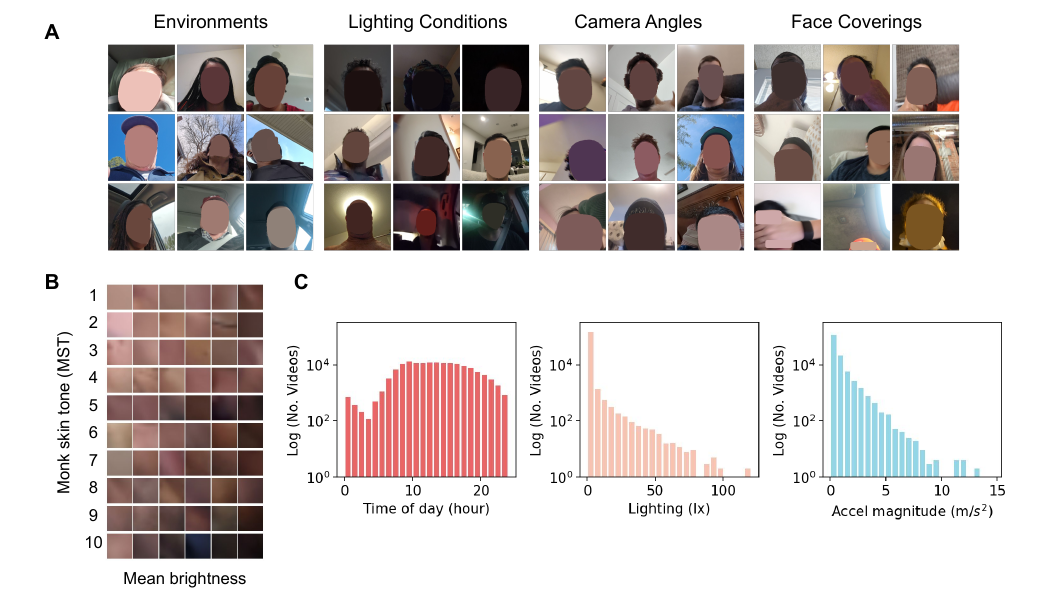}
	\caption{\textbf{Representative examples of the diversity of free-living data
	used to validate the PHRM.} (A) Illustrative examples of the variety of environments,
	lighting conditions, front-facing camera angles, and face obstructions for
	videos captured in the free-living conditions. (B) Examples of facial skin patches
	randomly sampled from video frames of the cheeks of participants across the full
	range of Monk skin tones (MST). Videos are sorted by mean brightness across
	columns and MST across rows (C) From left to right: histograms of the number of
	8-sec video clips by the hour of day, illuminance measured by the smartphone
	ambient light sensor, and the average magnitude of linear acceleration of the smartphone
	during the videos. }
	\label{fig:figure_2}
\end{figure*}

Participants uploaded 230.7 ± 172.2 face videos per day. These videos were recorded
passively throughout the day during normal personal phone use subsequent to a screen
unlock event. As expected, the unconstrained nature of free-living use and passive
recordings yielded videos with a diversity of environments,lighting conditions,
camera angles and face coverings (Fig.~\ref{fig:figure_2}A). These videos spanned
all hours of the day, and a wide range of lux and motion levels as measured by
the smartphone ambient light sensor and accelerometer, respectively (Fig.~\ref{fig:figure_2}C).
We randomly sampled skin patches from video frame crops of participant’s cheeks
to visualize the range of skin pigmentation under various lighting conditions
across the MST range (Fig.~\ref{fig:figure_2}B).

\begin{figure*}[!t]
	\centering
	\includegraphics[width=.9 \textwidth]{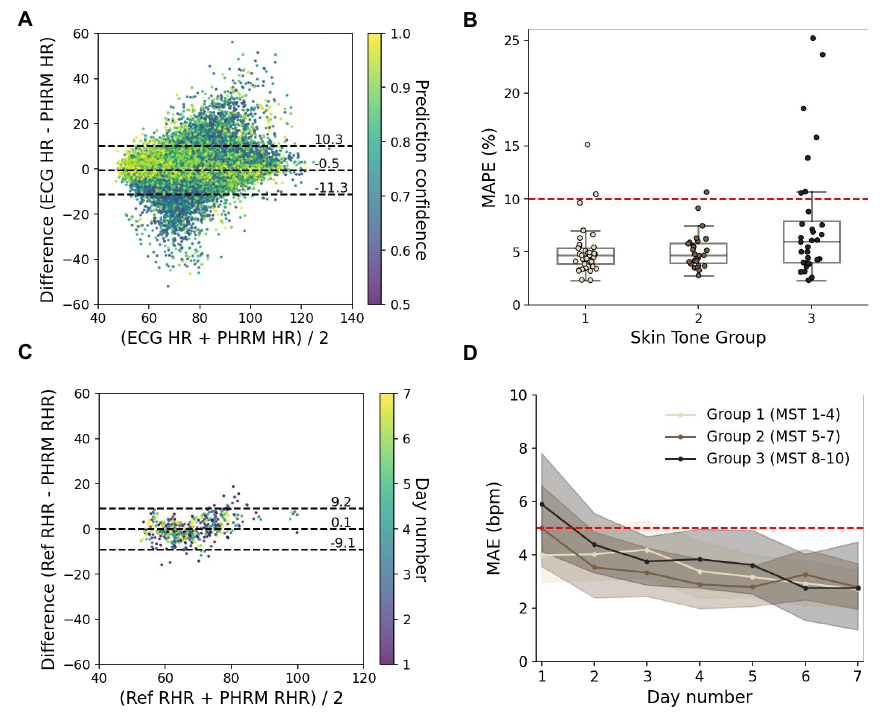}
	\caption{\textbf{Accuracy of passive HR and RHR measurements by the PHRM in
	free-living conditions.} (A) Bland Altman plot showing the agreement between PHRM-estimated
	HR values and the reference ECG measurements. Colors indicate the confidence
	level of PHRM predictions. Dashed lines show the bias, lower, and upper limits
	of agreement adjusted for repeated measurements with unequal numbers of replicates.
	(B) Boxplots showing the distribution of mean absolute percentage error (MAPE)
	values for individual participants, grouped by skin pigmentation. The box bounds the interquartile range (IQR) divided by the median, and whiskers extend to a maximum of 1.5 × IQR beyond the box. The red
	dashed line indicates the pre-specified accuracy target of MAPE $< 10$\%. (C)
	Bland Altman plot showing the agreement between PHRM-estimated daily RHR
	values and the reference wearable HR tracker measurements. Colors indicate the
	day number since the start of RHR predictions. Dashed lines show the bias, lower,
	and upper limits of agreement adjusted for repeated measurements with unequal
	numbers of replicates. (D) Mean absolute error (MAE) of PHRM-estimated RHR as a
	function of day number since the start of RHR predictions, grouped by skin
	pigmentation. Shaded areas indicate the 95\% confidence intervals. The red dashed
	line indicates the pre-specified accuracy target of MAE $< 5$ bpm. }
	\label{fig:figure_3}
\end{figure*}

\subsection{HR Measurements}
\paragraph{In-Laboratory Test Performance.}
We first explored how well smartphones measure HR in controlled conditions by
comparing PHRM predictions with HR measured by the reference ECG. In the prospective
laboratory study comprising 103 participants, we successfully obtained a valid
HR measurement (by gating on the confidence scores associated with the PHRM predictions,
see details in Methods) in 1360 out of 1750 face videos (77.7\%). The one participant
we did not obtain any valid HR measurements from was seated far from the camera,
resulting in a high failure rate (62.5\%) of detecting facial landmarks needed to
perform video stabilization. Compared to the reference ECG HR, the PHRM achieved
a mean absolute error (MAE) of 4.09 (95\%CI: 3.03, 5.33) and a mean absolute percentage
error (MAPE) of 5.65\% (95\%CI: 4.25, 7.29) at the participant level in the
overall study population (Table S~\ref{tab:hr3_result}). The MAPE values for all
five lighting conditions, and both at-rest and post-exercise conditions were
significantly lower than the pre-specified study target of 10\% (p $<0.001$),
indicating robustness across lighting and physiologic conditions. The PHRM measurement
success rate while participants were at-rest was 78.4\% (95\%CI: 76.3, 82.5), which
was higher than the 62.1\% (95\%CI: 56.6, 67.6) success rate post-exercise (Table
S~\ref{tab:hr3_rate}).

The Bland-Altman plot showed minimal bias (-0.7) and 95\% limits of agreement (LoAs),
adjusted for multiple measurements per participant, between -12.9 and 11.5 bpm (Fig.~\ref{fig:figure_2}).
The participant-level MAPE by skin tone groups was 3.81\% (95\%CI: 2.43, 5.94)
for Group 1, 4.43\% (95\%CI: 3.12, 6.06) for Group 2, and 8.93\% (95\%CI: 5.60, 12.60)
for Group 3; all were significantly $<10$\% (p < 0.025). The MAPE was highest for Group 3 under incandescent lighting.

\paragraph{Free-living Test Performance. }
Next, we evaluated the accuracy of passive HR measurements during smartphone use
in real-world conditions. In the free-living test set comprising 101
participants, we observed a lower HR measurement success rate of 43.1\% (68,307 valid
measurements out of the 158,471 face videos) compared to the laboratory study, as
expected given the measurements were in uncontrolled conditions without participants
being told to stay still during the duration of the measurement. The measurement
success rate decreased by skin tone group, ranging from 58\% in group 1, 45\% in
group 2, to 25\% in group 3. These videos were captured using 26 different
smartphone models. Compared to the reference ECG HR, the PHRM achieved a video-level
MAE of 3.59 (95\%CI: 3.33, 3.88) and MAPE of 4.83\% (95\%CI: 4.39, 5.28);
participant-level MAE was 4.58 (95\%CI: 4.14, 5.06) and MAPE was 6.09\% (95\%CI:
5.35, 6.87). Both video- and participant-level MAPEs were significantly < 10\% (p
$< 0.001$), indicating that the PHRM provided accurate passive HR measurements (Table~\ref{tab:main_freeliving_result}).

The participant-level MAPE by skin tone groups was 5.04\% (95\%CI: 4.36, 5.89)
for Group 1, 5.12\% (95\%CI: 4.51, 5.80) for Group 2, and 7.84\% (95\%CI: 6.25, 9.71)
for Group 3; all were significantly $< 10$\% (p $< 0.001$). The MAPE values for all
three skin tone groups were significantly $< 10$\% at both video and participant
level (p $< 0.001$ for all comparisons), demonstrating that the PHRM provided accurate
passive HR measurements for all skin pigmentation groups. The difference in
participant-level MAPE between skin pigmentation group 1 versus others, groups 2
versus others, and group 3 versus others was -1.64 percentage points (95\%CI: -2.96,
-0.37), -1.31 points (95\%CI: -2.51, -0.17), and 2.75 points (95\%CI: 1.04, 4.70),
respectively. This met the prespecified target of non-inferiority of $< 5$
percentage points, indicating that the PHRM provided equitable HR measurements for
all skin pigmentation groups. The Bland-Altman plot showed minimal bias (0.64)
and 95\% LoAs, adjusted for multiple measurements per participant, between -11.3
and 10.3 bpm (Fig.~\ref{fig:figure_3}A). Importantly, videos with lower errors
tended to have higher confidence scores, illustrating the effectiveness of the confidence-based
gating algorithm. Overall, our results indicate that PHRM was robust to diverse lighting,
physiologic and real-world conditions, and provided accurate and equitable HR measurements
across skin pigmentation groups.

\begin{table}[!t]
	\centering
	\caption{\textbf{Accuracy of passive heart rate (HR) and daily resting HR (RHR)
	measurements by the PHRM in free-living conditions across skin pigmentation
	groups.}}
	\resizebox{\textwidth}{!}{\tiny
\begin{tabular}{llllllll}
                    \toprule & \multirow{2}{*}{Group}                                             & \multicolumn{3}{l}{Video-level / Day-level Performance}                                                                                                                                 & \multicolumn{3}{l}{Participant-level Performance}                                                                                                                                        \\
                     &                                                                    & MAE                                   & RMSE                             & MAPE                        & MAE                                   & RMSE                              & MAPE                        \\ \midrule
\multirow{4}{*}{HR}  & \begin{tabular}[c]{@{}l@{}}Total\\ (n=101)\end{tabular}            & \begin{tabular}[c]{@{}l@{}}3.59\\ (3.33-3.88)\end{tabular}  & \begin{tabular}[c]{@{}l@{}}5.53\\ (5.10-6.03)\end{tabular} & \begin{tabular}[c]{@{}l@{}}4.83†\\ (4.41-5.29)\end{tabular}  & \begin{tabular}[c]{@{}l@{}}4.58\\ (4.15-5.06)\end{tabular}  & \begin{tabular}[c]{@{}l@{}}7.28\\ (6.53-8.06)\end{tabular}  & \begin{tabular}[c]{@{}l@{}}6.09†\\ (5.39-6.89)\end{tabular}  \\
                     & \begin{tabular}[c]{@{}l@{}}Group 1\\ (n=37; MST 1-4)\end{tabular}  & \begin{tabular}[c]{@{}l@{}}3.31\\ (2.99-3.64)\end{tabular}  & \begin{tabular}[c]{@{}l@{}}4.97\\ (4.48-5.47)\end{tabular} & \begin{tabular}[c]{@{}l@{}}4.55†*\\ (3.99-5.12)\end{tabular} & \begin{tabular}[c]{@{}l@{}}3.66\\ (3.29-4.11)\end{tabular}  & \begin{tabular}[c]{@{}l@{}}5.76\\ (5.05-6.49)\end{tabular}  & \begin{tabular}[c]{@{}l@{}}5.04†*\\ (4.36-5.89)\end{tabular} \\
                     & \begin{tabular}[c]{@{}l@{}}Group 2 \\ (n=27; MST 5-7)\end{tabular} & \begin{tabular}[c]{@{}l@{}}3.58\\ (3.22-4.06)\end{tabular}  & \begin{tabular}[c]{@{}l@{}}5.44\\ (4.74-6.28)\end{tabular} & \begin{tabular}[c]{@{}l@{}}4.56†*\\ (4.14-5.05)\end{tabular} & \begin{tabular}[c]{@{}l@{}}4.14\\ (3.61-4.75)\end{tabular}  & \begin{tabular}[c]{@{}l@{}}6.50\\ (5.49-7.49)\end{tabular}  & \begin{tabular}[c]{@{}l@{}}5.12†*\\ (4.51-5.82)\end{tabular} \\
                     & \begin{tabular}[c]{@{}l@{}}Group 3\\ (n=37; MST 8-10)\end{tabular} & \begin{tabular}[c]{@{}l@{}}4.52\\ (3.81-5.62)\end{tabular}  & \begin{tabular}[c]{@{}l@{}}7.16\\ (5.98-8.81)\end{tabular} & \begin{tabular}[c]{@{}l@{}}6.12†*\\ (4.87-7.84)\end{tabular} & \begin{tabular}[c]{@{}l@{}}5.81\\ (4.89-6.86)\end{tabular}  & \begin{tabular}[c]{@{}l@{}}8.99\\ (7.58-10.43)\end{tabular} & \begin{tabular}[c]{@{}l@{}}7.84†*\\ (6.24-9.70)\end{tabular} \\ \midrule
\multirow{4}{*}{RHR} & \begin{tabular}[c]{@{}l@{}}Total\\ (n=101)\end{tabular}            & \begin{tabular}[c]{@{}l@{}}3.62†\\ (3.13-4.10)\end{tabular} & \begin{tabular}[c]{@{}l@{}}4.65\\ (4.07-5.19)\end{tabular} & \begin{tabular}[c]{@{}l@{}}5.35\\ (4.65-5.90)\end{tabular}   & \begin{tabular}[c]{@{}l@{}}4.39†\\ (3.47-5.16)\end{tabular} & \begin{tabular}[c]{@{}l@{}}5.92\\ (4.49-6.98)\end{tabular}  & \begin{tabular}[c]{@{}l@{}}6.40\\ (5.18-7.42)\end{tabular}   \\
                     & \begin{tabular}[c]{@{}l@{}}Group 1\\ (n=37; MST 1-4)\end{tabular}  & \begin{tabular}[c]{@{}l@{}}3.44†\\ (2.67-4.34)\end{tabular} & \begin{tabular}[c]{@{}l@{}}4.38\\ (3.44-5.31)\end{tabular} & \begin{tabular}[c]{@{}l@{}}5.31\\ (4.24-6.43)\end{tabular}   & \begin{tabular}[c]{@{}l@{}}3.72†\\ (2.91-4.65)\end{tabular} & \begin{tabular}[c]{@{}l@{}}4.69\\ (3.80-5.56)\end{tabular}  & \begin{tabular}[c]{@{}l@{}}5.71\\ (4.45-7.11)\end{tabular}   \\
                     & \begin{tabular}[c]{@{}l@{}}Group 2 \\ (n=27; MST 5-7)\end{tabular} & \begin{tabular}[c]{@{}l@{}}3.47†\\ (2.88-4.06)\end{tabular} & \begin{tabular}[c]{@{}l@{}}4.40\\ (3.64-5.16)\end{tabular} & \begin{tabular}[c]{@{}l@{}}5.05\\ (4.14-5.84)\end{tabular}   & \begin{tabular}[c]{@{}l@{}}3.56†\\ (2.61-4.72)\end{tabular} & \begin{tabular}[c]{@{}l@{}}4.82\\ (3.45-6.39)\end{tabular}  & \begin{tabular}[c]{@{}l@{}}5.01\\ (3.83-6.32)\end{tabular}   \\
                     & \begin{tabular}[c]{@{}l@{}}Group 3\\ (n=37; MST 8-10)\end{tabular} & \begin{tabular}[c]{@{}l@{}}4.06†\\ (3.33-4.92)\end{tabular} & \begin{tabular}[c]{@{}l@{}}5.27\\ (4.29-6.23)\end{tabular} & \begin{tabular}[c]{@{}l@{}}5.74\\ (4.77-6.77)\end{tabular}   & \begin{tabular}[c]{@{}l@{}}5.86\\ (4.17-7.48)\end{tabular}  & \begin{tabular}[c]{@{}l@{}}7.76\\ (5.38-9.55)\end{tabular}  & \begin{tabular}[c]{@{}l@{}}8.40\\ (6.05-10.55)\end{tabular} \\ \bottomrule
\end{tabular}} \label{tab:main_freeliving_result}
	\parbox{\textwidth}{\raggedright \footnotesize Errors at the video/day level
	refers to the average error from all paired measurements. Errors at the participant
	level refers to the average of the mean error of individual participants.
	Errors in parentheses indicate 95\% confidence intervals. MAE, mean absolute error. MAPE, mean absolute percentage error. MST, Monk skin tone. RMSE, root mean squared error. \\†Met pre-specified
	targets: MAPE for HR $<10$\%, MAE for RHR $<5$ bpm (p $<0.05$) \\ *Met pre-specified
	non-inferiority HR target of $<5$ percentage points versus other skin tone groups}
\end{table}

\subsection{Daily RHR Measurements}

As our findings indicated that smartphones could measure HR passively in free-living
conditions, we tested our hypothesis that RHR could be estimated on a daily
basis using these intermittent HR measurements by comparing PHRM predictions
with daily RHR measured by the reference wearable HR tracker. Out of the 101 participants,
90 participants (89.1\%) had at least one or more days with $\geq 20$ valid HR measurements, which was the minimum number of measurements needed to
compute a valid daily RHR (Fig. S~\ref{fig:flow_chart}). Among these
participants, a valid daily RHR measurement was obtained in 504 out of 685 (73.6\%)
participant days. Compared to the reference wearable HR tracker, the PHRM
achieved a day-level and participant-level MAE of 3.62 bpm (95\%CI: 3.18, 4.09)
and 4.39 (95\%CI: 3.67, 5.17) for RHR measurements, respectively, in the overall
study population. Both MAE values were significantly lower than the pre-specified
study target of 5 bpm (p $< 0.001$), indicating that the PHRM provided accurate
daily RHR measurements measurements passively, during ordinary phone use. The
Bland-Altman plot showed minimal bias (0.1) and adjusted 95\% limits of agreement
were between -9.1 and 9.2 bpm.

The PHRM-derived daily RHR was very highly correlated with daily RHR from the
wearable HR tracker (r=0.87, p $< 0.001$); correlation with traditional methods
of measuring RHR using ECG in a supine (r=0.73, p $< 0.001$) and sitting position (r=0.74,
p $< 0.001$) was also high (Fig. S~\ref{fig:supp_rhr}). The intra-person
standard deviation for supine and sitting RHR was 5.08 bpm (95\%CI: 4.44, 5.78)
and 4.52 bpm (95\%CI: 3.95, 5.11), respectively. The coefficient of variation (CV)
for supine and sitting RHR was 7.85\% (95\%CI: 6.84, 9.01) and 6.68\% (95\%CI:
5.81, 7.64), respectively. We observed that reproducibility of PHRM-derived RHR
was higher compared to traditional methods with a significantly lower intra-person
standard deviation of 1.20 bpm (95\%CI: 1.06, 1.35) and coefficient of variation
(CV) of 1.77\% (95\%CI: 1.56, 1.99).

At the day level, MAE of daily RHR measurements by skin tone groups was 3.44 (95\%CI:
2.75, 4.62) for Group 1, 3.47 (95\%CI: 2.77, 4.25) for Group 2, and 4.06 (95\%CI:
3.23, 5.13) for Group 3; all were significantly $< 5$ bpm (p $< 0.001$). The participant-level
MAE by skin tone groups was 3.72 (95\%CI: 2.94, 4.57) for Group 1, 3.56 (95\%CI:
2.60, 4.77) for Group 2, and 5.86 (95\%CI: 4.18, 7.70) for Group 3; MAEs were
significantly $< 5$ bpm for Group 1 (p $< 0.005$) and 2 (p $< 0.001$) but did not
reach significance for Group 3 (p=0.32). However, we found that the MAE across
all skin tone groups decreased over time as the Kalman filter converged. From the
third day onwards, the MAE for Group 3 was significantly $< 5$ bpm. Visually, we
observed that the PHRM-derived RHR was able to capture similar trends as the
wearable HR tracker (Fig. S~\ref{fig:rhr_example}).

\paragraph{RHR Association with Health Factors}

Finally, we examined if the daily RHR estimated from smartphone use was
associated with well-established risk factors. We found that participants with a
higher PHRM-derived RHR, after adjusting for chronological age, sex and day of measurement,
were more likely to exhibit markers of poorer health, namely obesity and poor cardiovascular
fitness (Table~\ref{tab:gls_model}). In a generalized least square model of PHRM-derived
RHR, both higher BMI and lower cardiovascular fitness were independent predictors
($\beta$=1.92 ± 0.57 bpm, p $< 0.001$; and $\beta$=-1.90 ± 0.27 bpm, p $<0.001$
per 1SD increase of BMI and VO2max, respectively). Taken together, these results
demonstrate that the PHRM produced daily RHR estimates that were accurate and
associated with markers of health status.

\begin{table}[!t]
	\centering
	{ \caption{\textbf{Generalized least squares (GLS) model of PHRM-estimated daily resting heart rate (RHR).}} \resizebox{0.55 \textwidth}{!}{\tiny
\begin{tabular}{llllll}
\toprule
Covariate & \multicolumn{3}{l}{GLS model fit}                & \multicolumn{2}{l}{ANOVA test} \\
          & \multicolumn{3}{l}{Beta (bpm) per 1 SD increase} & X2      & p                    \\ \midrule
BMI       & \multicolumn{3}{l}{1.92 ± 0.57}                  & 11.4    & 0.0007               \\
VO2max    & \multicolumn{3}{l}{-1.90 ± 0.27}                 & 48.0    & \textless{}0.0001   \\
\bottomrule
\end{tabular}} \label{tab:gls_model} }
	\parbox{\textwidth}{\raggedright \footnotesize BMI, body mass index. SD, standard
	deviation. VO2max, maximal oxygen consumption.}
\end{table}
\section{Discussion}

To our knowledge, this is the first demonstration that smartphones can be used
to passively monitor both HR and RHR during normal phone usage in the real-world.
It is also the largest and most diverse prospective validation study of rPPG to date.
Importantly, smartphones were able to produce accurate HR measurements that meet
the American National Standards Institute (ANSI) and Consumer Technology Association
(CTA) standards for consumer HR monitors (MAPE $\leq 10$\%) \citep{Consumer_Technology_Association2018-fm}
across all skin pigmentation groups. We demonstrated that smartphone-based
rPPG meets the pre-specified non-inferiority target for performance across skin-tone
groups, which is critical for equitable measurement. We also found that
smartphones were able to produce accurate estimates of daily RHR; smartphone-derived RHR was associated with known risk factors
for cardiovascular disease.

In this study, we performed validation on 185,970 videos from 205 participants across
in-lab and free-living conditions. These participants spanned the entire MST and
captured race and ethnicity diversity in skin pigmentation relevant to the US
population according to the guidelines proposed by the FDA \citep{Center_for_Devices_and_Radiological_Health2023-hy}.
Existing rPPG datasets are much smaller in volume and were almost exclusively
collected in controlled (in-lab) settings, many capturing videos using digital single-lens
reflex (DSLR) cameras or devices from specialist imaging companies; the largest
of these studies include BP4D+ \citep{Zhang2016-yz} (1,400 videos from 140 participants),
VIPL-HR \citep{Niu2019-si} (2,378 videos from 107 participants), UCLA-rPPG \citep{Wang2022-vi}
(503 videos from 104 participants), and MMPD \citep{tang2023mmpd} (660 videos from
33 participants). Moreover, none of the rPPG datasets or prior studies meet the
diversity requirements proposed by the FDA in terms of distribution and range of
skin tones.

A recent study \citep{Savur2023-cr} showed promising feasibility of rPPG to provide
background measurements of HR using tablets provided to participants, achieving
95\% LoAs of between -12.4 and 14.6 bpm for HR accuracy at a measurement success
rate of 20\%. However, tablet usage was low and occurred under very limited conditions,
primarily while watching videos at home, which the authors attributed to the fact
that the tablet was a new device introduced to the participants. In contrast, our
study leveraged personal smartphones already in use by the participants to passively
capture measurements in an entirely free-living manner whenever they were naturally
interacting with their phones. Encouragingly, our PHRM system achieved better 95\%
LoAs between -11.3 and 10.3 bpm and a higher measurement success rate of 43\%
despite the more challenging environments. These results suggest that smartphones
could provide passive monitoring of HR throughout the day, providing within-day
temporal resolution that may be useful for tracking HR fluctuations due to
physical or mental stressors.

Previous state-of-the-art (SotA) rPPG methods have performed poorly for people
with darker skin pigmentation. For example, work evaluating prior SotA rPPG
methods on a geographically diverse population that included darker skin tones
found that the mean error was low (around 1 bpm) but the standard deviation was
high (11 bpm) \citep{Dasari2021-rg}. Previous meta-analysis of existing rPPG
datasets and methods reported that HR accuracy was significantly impacted for
the darkest skin pigmentation with the MAE increasing from around 3.4 bpm for
Fitzpatrick skin types I-V to 13.6 bpm in skin type VI \citep{Nowara2020-fy}.
This highlights the importance of evaluating HR performance on the darkest skin
pigmentations (Type VI) separately, and not in combination with Type V which is common
practice in the rPPG field. To address the previous SotA underperformance on
darker skin tones, we deliberately devoted approximately one third of our study population
to participants of the darkest skin pigmentations (Fitzpatrick Type VI or MST 8-10)
during recruitment. This enabled us to train rPPG models that were less
susceptible to skin pigmentation bias. In this work, the PHRM system achieved an
MAE of 5.2 and 5.81 bpm in the darkest skin pigmentations (Fitzpatrick Type VI or
MST 8-10) in controlled and free-living environments, respectively. Our findings
indicate that smartphone-based rPPG can provide accurate and equitable HR
measurements across all skin pigmentation groups.

Another important finding of this work is that smartphones can passively
estimate RHR, which has not been previously demonstrated. Notably, we found that
PHRM-based RHR had a lower intra-person standard deviation and CV compared to traditional
methods, indicating it was more consistent and reproducible. Presumably, this is
because many observations of HR throughout the day capture more consistent RHR
values than a single measurement in the supine or sitting position \citep{Dunn2021-wy}.
Smartphone-based RHR was significantly associated with known risk factors for
cardiovascular health that influence clinic-based RHR, including obesity \citep{Rogowski2009-wx, Itagi2020-bb, Martins2003-eo},
and low VO2max \citep{Nauman2011-wt}. The results of our system on this study population
demonstrate its potential clinical significance and suggest the possibility of using
it to ambiently monitor elements of health status via the indicator of RHR. This
opens up the possibility of automatically collecting longitudinal RHR data
across weeks, months, seasons, and years that may provide valuable health information.

Our work has some limitations. There were a few participants with high MAPEs. From visual inspection of the videos from these participants, we observed frequent head motion and talking. The measurement success rate was lower in the
darkest skin pigmentation group after automated confidence-based gating. This is
likely related to the signal-to-noise (SNR) of the pulse signal captured in the videos,
which has been reported to decrease with darker skin pigmentation \citep{De_Haan2013-up, wang2014exploiting}.
Lower SNR has been attributed to the increased melanin content in darker skin
pigmentation that limits the amount of light entering the deeper skin layers
with pulsatile blood vessels and absorbs a portion of diffuse reflections
carrying pulsatile information, while the specular reflections are not reduced \citep{De_Haan2013-up, wang2014exploiting}.
Additionally, we cannot rule out other factors that may differ between the skin
tone groups, such as participant physical activity levels, phone device hardware,
and environment or lighting conditions. A potential broader mitigation would be
to increase the number of measurement attempts when low SNR is encountered. Relatedly,
under controlled conditions, we also observed that the MAPEs were highest for the
darkest skin tone group under incandescent lighting. This could be due to the
fact that incandescent lighting contains much less spectral power in the green
wavelengths \citep{Abdel-Rahman2017-dp}, which is optimal for blood absorption
and hence SNR of the pulse signal. One approach is to investigate optimizing the
camera exposure settings to boost the SNR, which might improve both measurement
success rate and accuracy under such lighting. Globally, incandescent lighting
is increasingly uncommon as there are ongoing efforts in multiple countries to
phase out incandescent lightbulbs to promote energy efficiency \citep{edge2008light}.

In this research study, we did not account for constraints on battery
consumption in favor of collecting more data and thus opportunistically
collected videos any time the phone was unlocked. To minimize smartphone power use,
future work is needed to identify the most opportune conditions likely to yield
an accurate HR measurement before activating the camera. This would also increase
the measurement success rate. Further improvements to the smartphone-based RHR
algorithm might also be possible, for example, by considering the time of each
HR measurement to account for circadian rhythms, or by using the accelerometer data
to identify HR measurements taken after a sufficiently long rest period.

There are important privacy concerns to be considered for respectful use of this
technology. Smartphone users should be asked to grant explicit informed consent
before enabling passive video-based HR measurement. In our studies, videos were
taken of consented participants and first saved locally on their device. The participants
then reviewed the videos and manually authorized upload for research use. They
were instructed not to upload any videos containing sensitive content or faces
other than their own. The PHRM system was designed such that it could be run locally
on a smartphone’s processors, which enables the videos to be processed on-device.
Such a system could be implemented within a protected on-device environment isolated
from unauthorized access, such as Android’s Trusted Execution Environment, to ensure
the video images remain secure during execution. In addition, implementation of such
a system could make HR measurement contingent on successful face authentication,
mitigating measurements of other individuals and incorrect HR data attribution.

In conclusion, we developed and validated a system for passive measurement of HR
and daily RHR during normal phone use that performs accurately across all skin pigmentation
groups. This advancement in the state-of-the-art for rPPG methods presents a
promising approach to improve equitable access to the benefits of heart health
tracking by widening its availability to everyone who has a smartphone.
\section{Methods}

\subsection{Studies}
Between October 2020 to March 2024, we conducted five independent, prospective laboratory
studies and a prospective free-living study to obtain datasets to develop and validate
the PHRM. All study protocols were approved by an Institutional Review Board (Quorum
now known as Advarra, Advarra, Columbia MD and WCG, Puyallup WA). We obtained
informed consent from all participants, and the study was conducted in accordance
with the principles of the Declaration of Helsinki.

In the laboratory validation studies, we objectively measured skin tone from each
participant by using a RM200QC spectrocolorimeter (Pantone LLC, Carlstadt NJ) to image
the skin of the cheeks and forehead. For the free-living study, since it was entirely
remote with no in-person component, we provided participants with a visual
representation of the Monk skin tone (MST) \citep{monk2019monk} to self-assess their
skin tone.

\paragraph{Reference Measurements}
To validate HR measurements of the PHRM in laboratory settings, we used ECG recorded
by the BIOPAC MP160 system (BIOPAC, Inc., Santa Barbara, CA) as the reference ground
truth. We used a custom LabVIEW (National Instruments, Woburn, MA) application
to record 3-lead ECG signals from electrodes placed on study participants’ upper
chests (or upper arms) and lower abdomens.

For validating HR measurements of the PHRM in real-world, free-living conditions,
we used the Polar H10 ECG chest strap (Polar, Kempele Finland). The Polar H10
has been validated to provide accurate HR measurements during physical activity
\citep{Gilgen-Ammann2019-pc, Pasadyn2019-gs}. Participants were instructed to
put the chest strap on every morning and to wear it for at least 7 hours each
day, except during showers or sleep.

Since aggregating multiple watch HR measurements provides more consistent RHR values
than spot measurements in a supine or sitting position \citep{Dunn2021-wy}, we chose
to use the daily RHR from the Fitbit Charge 6 (Google, Mountain View, CA) as our
primary reference for RHR. Daily RHR produced by Fitbit devices is computed by
combining multiple HR measurements across “at rest” periods throughout the day, where
the on-device accelerometer has determined that the person is at rest, and has
not recently been moving. If available, sleeping HR is also used to improve the
daily RHR estimate. The Fitbit Daily RHR has been shown to be closest to RHR measurements
taken lying down immediately upon wake \citep{Russell2019-wh}. In addition,
participants were instructed to perform two traditional RHR measurements first
thing in the morning, before eating, drinking, exercising, or showering. After putting on the ECG chest strap, they laid in a supine position for 6 minutes. Next, they sat
still for another 6 minutes. Supine and sitting RHR measurements were computed as
the minimum HR from the ECG recordings and served as secondary references. HR has
been found to stabilize in most subjects after 4 minutes of inactivity \citep{Speed2023-bm}.

\paragraph{Time Synchronization}
To synchronize the clocks across all the study devices during the free-living study, participants performed a
daily routine comprising a series of three jumps. We instructed participants to
stand still with their hands placed in front of the chest. They held their smartphone
enrolled in the study in their dominant hand; the wearable HR tracker was placed
on the off-hand. To perform the series of jumps, we asked participants to start
a timer, and complete the following sequence: standing still for one minute, three
jumps spaced by 10 seconds, followed by standing still for another 10 seconds. We
aligned the timestamps of the smartphone and ECG chest strap by maximizing the cross
correlation between their respective accelerometer signals after resampling the signals
to 60 Hz.

\subsection{PHRM-HR Module}

The PHRM-HR module is the component of our algorithm that predicts a HR
measurement. The PHRM-HR module processes an 8-second video input to predict HR.
In this section, we describe the preprocessing pipeline applied to raw video,
the deep network used for HR extraction, and the confidence-gating algorithm.

\paragraph{Developmental Metric.}
To assess and optimize each sub-component of the PHRM-HR module, we used the tuning
dataset and computed the root mean square error (RMSE) after excluding the bottom
20\% of videos with the lowest confidence scores. This approach was chosen to improve
robustness against outliers and maintain sensitivity to extreme HR predictions.

\paragraph{Video Preprocessing}
The video preprocessing pipeline consists of five key steps to prepare the video
data for HR extraction. Each step is briefly outlined in the following paragraphs,
with further details provided in Appendix~\ref{app:hr_ablation_video_preprocessing}.
\begin{itemize}
	\item \textbf{Stabilization.} Video stabilization ensures consistent facial
		alignment across frames. To achieve this, facial landmarks were first
		detected using the Android face-based augmented reality AR model \citep{Unknown2024-gw},
		and their positions were averaged to compute a centroid point. Then, an
		affine transformation was applied based on this centroid to stabilize each
		frame.

	\item \textbf{Linear interpolation.} Due to mobile I/O constraints, videos were
		stored at varying frame rates. All videos were standardized to 15 FPS for
		consistency. For videos not recorded at 15 FPS, each pixel value across the
		temporal dimension was interpolated linearly to achieve the target frame
		rate.

	\item \textbf{Face cropping.} To reduce background noise, the Android face-based
		AR model was used to detect the bounding box of the face in each frame. Then,
		we calculated the minimal bounding box covering the face across all frames,
		with an additional 20\% margin in each dimension to include the ears and neck.

	\item \textbf{Resizing.} To ensure mobile compatibility with respect to
		limited computation, all video frames were resized to 32x32 pixels. In our
		experiments, we found this resolution was the smallest resolution that still
		retained HR information and anti-aliased resampling with area interpolation proved
		to be the most effective method for preserving HR data at this resolution.

	\item \textbf{Frame differencing.} To highlight pixel value changes associated
		with physiological signals, we computed the difference between consecutive frames,
		similar to calculating the first derivative of a physiological signal.
\end{itemize}

\paragraph{PHRM-HR Network}
We trained a deep learning model to extract HR from video data that was preprocessed
as described above. Specifically, the network takes 8-second videos at 15 FPS (a
total of 120 frames) as input and outputs a continuous HR value between 40 and 180
BPM. In this section, we describe the network architecture and training process of
our model. Ablation experiments were conducted to evaluate each component, with further
details provided in Appendix~\ref{app:hr_ablation_network}.
\begin{itemize}
	\item \textbf{Network backbone.} We used a temporal shift (TS) convolutional
		network \citep{Liu2020-tg} to process each frame, generating a pseudo-PPG
		signal along the temporal axis. The network consists of five TS blocks, with
		each block consisting of a 2D convolution layer, ReLU activation, temporal
		channel shifting, and layer normalization across the channel dimension. The
		number of channels per block was set to 4, 16, 32, 64, and 128, respectively,
		with a fixed kernel size of 3. After passing through the TS blocks, average pooling
		was applied across all channels for each frame, followed by a temporally
		shared linear layer to produce the pseudo-PPG signal.

	\item \textbf{HR output head.} To extract HR from the pseudo-PPG, we first applied
		a Fast Fourier Transform to convert the signal into the frequency domain.
		The frequencies were then bucketized into 1 Hz bins, with the corresponding magnitudes
		treated as categorical logits. A softmax function was used to convert these
		logits into HR probabilities. In evaluation, the predicted HR was computed
		as a weighted sum of the probabilities.

	\item \textbf{Loss function.} We reformulated the HR regression problem as a
		classification task \citep{Lathuiliere2020-dv}. The ground truth HR value
		was bucketized into a one-hot vector, corresponding to the relevant HR bins,
		with a bin size of 1 Hz. Focal loss \citep{Lathuiliere2020-dv, Ross2017-vb} was then applied to the HR probabilities
		and the bucketized HR values, with $\alpha$=5 and $\gamma$=6 tuned during
		training to optimize network performance.

	\item \textbf{Optimization.} We trained the model using the Adam optimizer
		\citep{Ross2017-vb, Kingma2014-ht, Gotmare2018-ze} with a learning rate warm-up
		and a cosine decay scheduler48–51. Specifically, the model was trained for 20,000
		steps, with 3,000 steps allocated for warm-up. Cosine decay was applied for 5,000
		steps with a decay rate of 0.95. The learning rate and batch size were
		determined through a hyper-parameter search on the tuning dataset.

	\item \textbf{Data augmentation.} We applied two sets of augmentations to
		enhance model generalization. The first set consisted of spatial augmentations
		\citep{Liu2024-oc}, including up/down and left/right flipping, rotation, and
		cropping and resizing, applied in a consistent manner across the temporal dimension.
		The second set introduced temporal noise through speed augmentation \citep{Yang2022-jg},
		where video playback was randomly adjusted by a factor between 0.8 and 1.2. Speed
		augmentation was implemented using the same linear interpolation method described
		in Video Preprocessing. Each augmentation was applied independently, with probabilities
		and parameters optimized through hyper-parameter search.

	\item \textbf{Hyper-parameter search.} Given the extensive hyper-parameter
		search space (learning rate, batch size, augmentation probabilities, etc.), we
		used a Gaussian process-based search tool to find the optimal parameter combination
		\citep{Golovin2017-pn}. We set the search to 400 runs, using the post-confidence gated RMSE on validation as the evaluation metric, and the searched
		hyper-parameters are presented in Table S~\ref{tab:hyper-parameter}.

	\item \textbf{Ensembling.} We selected the top five models from the hyper-parameter
		search based on their validation performance and averaged the per-model HR
		prediction to generate the final HR prediction.
\end{itemize}

\paragraph{Confidence Gating}
Due to the unconstrained environment of normal phone use in real-world settings,
it was necessary to apply a gating criterion to discard face videos that were too
noisy for reliable HR estimation, such as those with no face present, face coverings
or excessive movement. We found that the deep learning model’s confidence of a
HR prediction was an effective metric for this purpose and derived an optimal
threshold using the tuning dataset. Specifically, we use the negative entropy of
the HR probabilities generated by the PHRM-HR module, where a higher negative entropy
indicates a higher confidence. In the Supplementary Materials, we compared two alternatives,
including pseudo-PPG SNR and the maximum HR probability, and found negative entropy
to be the most effective for determining valid HR measurements.

Another key aspect of gating is determining the cutoff, which defines the
threshold below which videos are filtered. To ensure this cutoff is accurate and
equitable across different skin tones, we applied two rules when searching for
the cutoff in the free-living tune dataset. First, the overall MAPE for each
skin tone group had to be $<8$\%. Second, the MAPE gap between any two skin tone
groups had to be $<3$\%. The search process is detailed in Appendix~\ref{app:hr_ablation_video_gating}.

\subsection{PHRM-RHR Model}
For the daily RHR algorithm, we maintained simplicity to enhance its
generalization. First, we aggregated the valid HR measurements across a single
day by computing the 10th percentile value and applying a bias correction factor
(which is a constant across all participants). We performed a grid search for the
optimal percentile and bias correction values using the tune dataset. Next, we applied
a Kalman filter to refine the RHR prediction from noisy estimates. Using the
tune dataset, we also identified the minimum number of valid HR measurements needed
in a day to provide a valid RHR estimate.

\subsection{Statistical Analysis}
For HR measurements, we established a pre-determined accuracy target of mean
absolute percentage error (MAPE) $<10$\% in accordance with the ANSI/CTA standards
for consumer HR monitors \citep{Consumer_Technology_Association2018-fm}, which is
based on the ANSI/AAMI standard for HR accuracy for ECG monitors \citep{Association-for-the-Advancement-of-Medical-Instrumentation-AAMI-2002-tm}.
Measurements were paired observations: PHRM-estimated HR and reference HR from
ECG. A paired measurement was dropped if the PHRM did not produce a valid HR measurement,
i.e. the confidence of PHRM predictions was lower than the gating cutoff. Each
participant contributed multiple 8-second videos for HR measurements. We computed
MAPE at the video level as the mean value for all absolute percent error values from
each paired measurement and used bootstrapping at the participant level to obtain
the 95\% confidence intervals (CI). Since each participant contributed a different
number of videos, we also computed MAPE at the participant level as the mean
value for the MAPEs of individual participants, which we visualized using boxplots.
Since the errors were not normally distributed based on a Shapiro-Wilk test, we determined
if the participant-level MAPE values were significantly $<10$\% if p $<0.05$
using the Wilcoxon sign-rank test. We determined non-inferiority if the upper limit
of the 95\% CI around the difference in MAPE across participants in any of the
three skin pigmentation groups compared to that across participants in the other
two skin pigmentation groups was less than a pre-specified 5 percentage points. Bland–Altman
plots were used to visualize the agreement between the estimated values and the reference
measurements; limits of agreement were adjusted for repeated measurements with
unequal numbers of replicates \citep{Bland1999-ai}.

For daily RHR measurements, we adopted a pre-specified accuracy target of mean
absolute error $< 5$ bpm, which provided a stricter requirement than MAPE $<10$\%.
Similar to above, measurements were paired observations: PHRM-estimated daily RHR
and reference daily RHR from the wearable HR tracker. For the PHRM to yield a
valid daily RHR estimate, a minimum of 20 valid HR measurements on that day was needed.
Each participant contributed multiple days for RHR measurements. We computed MAE
at the day level as the mean value for all absolute error values from each paired
measurement and used cluster bootstrapping to obtain the 95\% confidence
intervals (CI). To account for multiple observations, we also computed MAE at
the participant level as the mean value for the MAEs of individual participants,
which we visualized using boxplots. Since the Kalman filter in the PHRM-RHR
module takes time to converge, we also computed MAE for each day since the start
of PHRM-RHR predictions to evaluate the MAE over time. We determined if MAE
values were significantly $< 5$ bpm if p $<0.05$ using the Wilcoxon sign-rank
test. Bland–Altman plots were used to visualize the agreement between the estimated
values and the reference measurements; limits of agreement were adjusted for
repeated measurements with unequal numbers of replicates \citep{Bland1999-ai}.

Associations between PHRM-estimated daily RHR and known risk factors were evaluated
using a generalized least square model accounting for the correlation structure present
across the serial estimates of RHR while also adjusting for the clusters within participants
\citep{Harrell2015-rz}. Ground truth measurements and associated estimates of RHR
were collected daily for each participant over the course of the study.
Therefore, statistical models of RHR have to: 1) account for the serial correlation
existing among repeated measurements and 2) adjust for the presence of clusters
within subjects. Given its flexibility, we decided to follow a generalized least
squares (GLS) modeling approach \citep{strategies2001applications}. First, we evaluated
several possible forms for the correlation structure across days of measurement given
a subject cluster, and ultimately selected the “spherical correlation” form, as it
was most suited to our data in terms of Akaike Information Criterion. We then
fitted a GLS model of the PHRM-estimated RHR using BMI and VO2max as covariates
and adjusting for age, sex, age*sex and day of measurement. The analyses were conducted
in R \textit{v4.41} leveraging the rms \textit{v6.8} and nlme \textit{v3.1}
packages.
\subsection*{Acknowledgements}
We thank Nikola Teslovich, Alex Mun, Jonathan Hsu, Xiaoxia Zhang, Derrick Vickers, Sam Mravca, Tracy Giest, Jiening Zhan for their support of this work.

\subsubsection*{Competing interests}
This study was funded by Alphabet Inc and/or a subsidiary thereof (‘Alphabet’). All authors are employees of Alphabet and may own stock as part of the standard compensation.

\subsubsection*{Data availability}
The data that support the findings of this study are not openly available due to reasons of study participant privacy. The informed consent agreements do not allow for sharing of participant-level data.

\subsubsection*{Code Availability}
A pseudocode implementation of the algorithms is available in the Supplementary Materials.

\bibliography{main}

\clearpage
\appendix

\setcounter{figure}{0}    
\setcounter{table}{0}    

\section*{\LARGE Supplementary Material}

\startcontents[appendices]

%\clearpage
\section{Additional Information for Free-Living Protocol}
\label{app:freeliving_protocal}

\begin{figure*}[h!]
    \centering
    \includegraphics[angle=0, width=1 \textwidth]{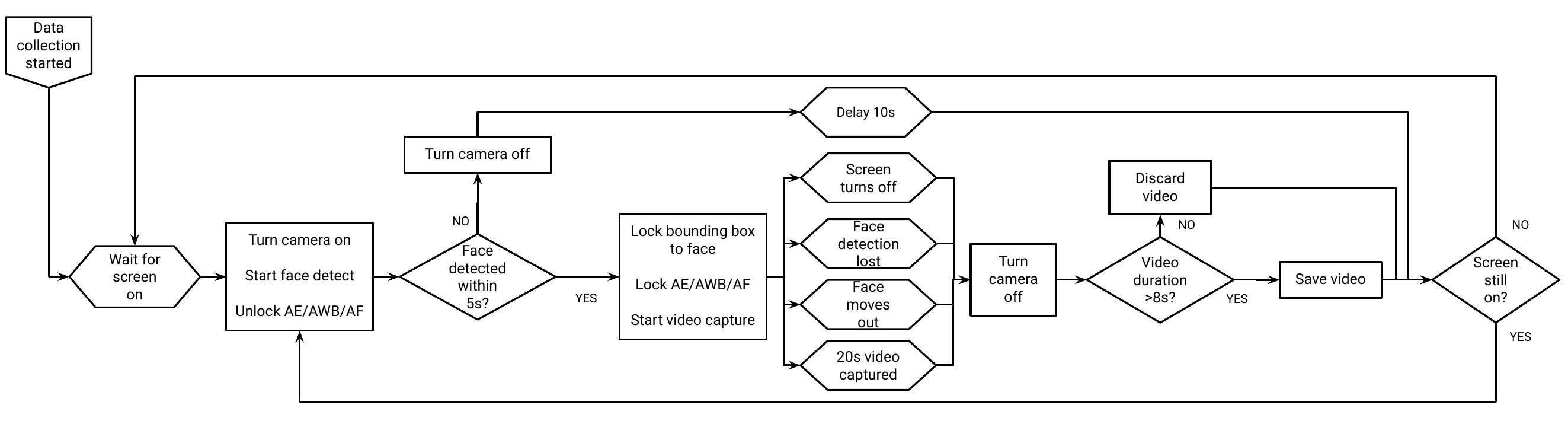}
    \caption{\textbf{Sequence of events in an ambient video recording.} Upon screen unlock, the front-facing camera would start at VGA resolution (640x480 px), 15 FPS, and with the phone’s default 3A settings (autoexposure, autofocus, auto-white balance) enabled. If a face was detected within five seconds, the camera would lock the current 3A settings and begin recording frames. Recorded frames were cropped to a stationary bounding box set by the initial position of the face, and saved as Motion JPEG (M-JPEG) at maximum quality to avoid inter-frame compression. The recording would automatically end after 20 seconds had elapsed, if the face moved out of the bounding box, or if the screen was turned off. Clips shorter than 8 seconds were discarded. If the screen was still on, this sequence would restart with the camera reset to its initial settings. This sequence could repeat up to 30 times per screen unlock, for a maximum 10 minutes of video.
    }
    \label{fig:video_protocal} 
\end{figure*}

This section details the protocol used for collecting free-living data, aimed at capturing face videos and ground truth measurements in natural, real-world settings. Participants were instructed to follow a specific morning routine, including capturing RHR using a Fitbit and Polar chest strap, as well as aligning accelerometer signals across devices. Throughout the day, video clips were recorded periodically as participants used their phones. To ensure privacy and transparency, several safeguards were implemented, including participant review of video clips, restrictions on recording conditions, and cropping of videos to only show the participant's face. Fig.~\ref{fig:video_protocal} illustrates the sequence of events during the ambient video recording process.

\paragraph{Participant Instruction}
Free-living data collection aimed to collect face videos and ground truth under natural, real-world conditions. Participants were instructed to adhere to the following protocol each morning:
\begin{itemize}
    \item Immediately after waking up, begin data collection on their phone, Fitbit, and Polar chest strap
    \item Lie still for 6 minutes, then sit still for 6 minutes to measure resting heart rate on the Fitbit / Polar
    \item Jump three times while holding the phone and wearing the Fitbit / Polar, thus creating accelerometer signals across all three devices to be used for timestamp alignment
\end{itemize}

Participants then used their phone throughout the day as they normally would. Brief video clips would be periodically recorded from the front-facing camera when the phone was in use.

\paragraph{Transparency \& privacy}
Given the intrusive nature of this data collection, several precautions were implemented to avoid inadvertently collecting sensitive material and to be transparent about the collected data:
\begin{itemize}
    \item Video data never left the phone until it was manually reviewed and approved by the participant. Participants were instructed to view each video clip and delete clips containing other people or any sensitive subjects.
    \item Recording could only occur within 10 minutes of the phone being unlocked and only while a face was visible. Recording immediately stopped whenever the phone was locked or if no face was in frame. This was intended to reduce the chance of recording someone other than the participant.
    \item Videos were cropped automatically (based on the Android AR model) to only the participant’s face to avoid recording the background environment.
    \item On-screen indicators were displayed whenever a video recording was in progress.
\end{itemize}

\clearpage
\newpage
\section{Additional Results in Laboratory Setting}
\label{app:inlab_results}

\begin{figure*}[h!]
    \centering
    \includegraphics[width=1 \textwidth]{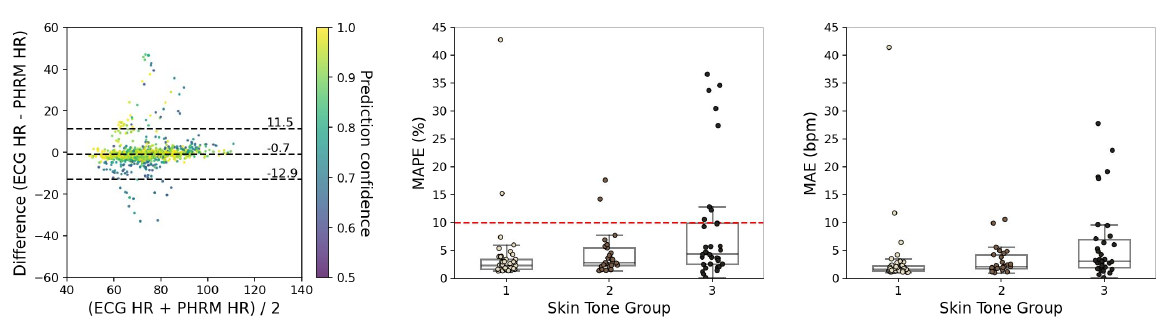}
    \caption{\textbf{Accuracy of HR measurements by the PHRM in laboratory settings.} (A) Bland Altman plot showing the agreement between PHRM-estimated HR values and the reference ECG measurements. Colors indicate the confidence level of PHRM predictions. Dashed lines show the bias, lower, and upper limits of agreement adjusted for repeated measurements with unequal numbers of replicates. (B) Boxplots showing the distribution of mean absolute percentage error (MAPE) values for individual participants, grouped by skin pigmentation. The box bounds the interquartile range (IQR) divided by the median, and whiskers extend to a maximum of 1.5 × IQR beyond the box. The red dashed line indicates the pre-specified accuracy target of MAPE $<10$\%.   }
    \label{fig:hr3_result} 
\end{figure*}

In this section, we present tables and figures that summarize the performance of the PHRM in laboratory conditions, evaluating its accuracy and success rates under diverse lighting scenarios. Table~\ref{tab:hr3_result} shows the accuracy of HR measurements, providing detailed results on how well the PHRM matches reference measurements across varying conditions. Meanwhile, Table~\ref{tab:hr3_rate} focuses on the success rate of measurements, indicating how often the PHRM successfully captured valid HR readings. In addition, Fig.~\ref{fig:hr3_result} visualizes these findings through a Bland-Altman plot and boxplots. The Bland-Altman plot illustrates the agreement between PHRM-estimated HR values and reference ECG measurements, with color coding for prediction confidence levels, while the boxplots highlight the distribution of MAPE and MAE across participants, categorized by skin pigmentation.

\begin{table}[H]
    \centering
    \caption{\textbf{Accuracy of heart rate (HR) measurements by the PHRM in diverse laboratory conditions and lighting.}}
    \resizebox{\textwidth}{!}{\begin{tabular}{lllllllllll}
                                                      \toprule & \multirow{2}{*}{Group}                                      & \multirow{2}{*}{All}                                           & \multicolumn{2}{l}{Condition}                                                                                                 &  & \multicolumn{5}{l}{Lighting (at-rest)}                                                                                                                                                                                                                                                                                            \\
                                                       &                                                             &                                                                & At-Rest                                                       & Post-Exercise                                                 &  & Fluorescent                                                    & Incandescent                                                   & Natural                                                        & Dim LED                                                       & Normal LED                                                     \\ \midrule
\multirow{4}{*}{MAE}             & \begin{tabular}[c]{@{}l@{}}Total\\ (n=103)\end{tabular}     & \begin{tabular}[c]{@{}l@{}}4.09\\ (3.03 - 5.33)\end{tabular}   & \begin{tabular}[c]{@{}l@{}}4.28\\ (3.19 - 5.59)\end{tabular}  & \begin{tabular}[c]{@{}l@{}}2.16\\ (1.71 - 2.70)\end{tabular}  &  & \begin{tabular}[c]{@{}l@{}}3.69\\ (2.51 - 5.15)\end{tabular}   & \begin{tabular}[c]{@{}l@{}}4.06\\ (2.85 - 5.53)\end{tabular}   & \begin{tabular}[c]{@{}l@{}}3.72\\ (2.41 - 5.39)\end{tabular}   & \begin{tabular}[c]{@{}l@{}}3.62\\ (2.45 - 5.11)\end{tabular}  & \begin{tabular}[c]{@{}l@{}}4.24\\ (2.88 - 5.90)\end{tabular}   \\
                                                       & \begin{tabular}[c]{@{}l@{}}Group 1\\ (n=44)\end{tabular}    & \begin{tabular}[c]{@{}l@{}}3.00\\ (1.74 - 5.09)\end{tabular}   & \begin{tabular}[c]{@{}l@{}}3.04\\ (1.75 - 5.20)\end{tabular}  & \begin{tabular}[c]{@{}l@{}}1.85\\ (1.48 - 2.29)\end{tabular}  &  & \begin{tabular}[c]{@{}l@{}}3.42\\ (1.73 - 5.88)\end{tabular}   & \begin{tabular}[c]{@{}l@{}}2.80\\ (1.56 - 4.83)\end{tabular}   & \begin{tabular}[c]{@{}l@{}}2.87\\ (1.53 - 5.29)\end{tabular}   & \begin{tabular}[c]{@{}l@{}}3.22\\ (1.69 - 5.75)\end{tabular}  & \begin{tabular}[c]{@{}l@{}}3.16\\ (1.63 - 5.36)\end{tabular}   \\
                                                       & \begin{tabular}[c]{@{}l@{}}Group 2\\ (n=25)\end{tabular}    & \begin{tabular}[c]{@{}l@{}}3.16\\ (2.28 - 4.23)\end{tabular}   & \begin{tabular}[c]{@{}l@{}}3.31\\ (2.37 - 4.43)\end{tabular}  & \begin{tabular}[c]{@{}l@{}}1.93\\ (1.16 - 2.98)\end{tabular}  &  & \begin{tabular}[c]{@{}l@{}}2.62\\ (1.69 - 3.63)\end{tabular}   & \begin{tabular}[c]{@{}l@{}}4.02\\ (2.46 - 5.95)\end{tabular}   & \begin{tabular}[c]{@{}l@{}}3.73\\ (1.84 - 6.77)\end{tabular}   & \begin{tabular}[c]{@{}l@{}}2.40\\ (1.70 - 3.35)\end{tabular}  & \begin{tabular}[c]{@{}l@{}}3.24\\ (1.87 - 4.98)\end{tabular}   \\
                                                       & \begin{tabular}[c]{@{}l@{}}Skintone 3\\ (n=34)\end{tabular} & \begin{tabular}[c]{@{}l@{}}6.17\\ (4.03 - 8.62)\end{tabular}   & \begin{tabular}[c]{@{}l@{}}6.70\\ (4.32 - 9.43)\end{tabular}  & \begin{tabular}[c]{@{}l@{}}3.09\\ (1.80 - 4.78)\end{tabular}  &  & \begin{tabular}[c]{@{}l@{}}5.25\\ (2.74 - 8.33)\end{tabular}   & \begin{tabular}[c]{@{}l@{}}7.05\\ (3.72 - 11.26)\end{tabular}  & \begin{tabular}[c]{@{}l@{}}5.96\\ (3.18 - 9.40)\end{tabular}   & \begin{tabular}[c]{@{}l@{}}5.74\\ (3.03 - 9.17)\end{tabular}  & \begin{tabular}[c]{@{}l@{}}7.53\\ (3.75 - 12.15)\end{tabular}  \\ \midrule
\multirow{4}{*}{MAPE} & \begin{tabular}[c]{@{}l@{}}Total\\ (n=103)\end{tabular}     & \begin{tabular}[c]{@{}l@{}}5.65†\\ (4.25 - 7.29)\end{tabular}  & \begin{tabular}[c]{@{}l@{}}6.01†\\ (4.43 - 7.75)\end{tabular} & \begin{tabular}[c]{@{}l@{}}2.74†\\ (2.11 - 3.52)\end{tabular} &  & \begin{tabular}[c]{@{}l@{}}4.95†\\ (3.54 - 6.58)\end{tabular}  & \begin{tabular}[c]{@{}l@{}}5.90†\\ (4.08 - 8.06)\end{tabular}  & \begin{tabular}[c]{@{}l@{}}4.88†\\ (3.42 - 6.70)\end{tabular}  & \begin{tabular}[c]{@{}l@{}}5.15†\\ (3.52 - 7.23)\end{tabular} & \begin{tabular}[c]{@{}l@{}}5.75†\\ (4.02 - 7.81)\end{tabular}  \\
                                                       & \begin{tabular}[c]{@{}l@{}}Group 1\\ (n=44)\end{tabular}    & \begin{tabular}[c]{@{}l@{}}3.81†\\ (2.43 - 5.94)\end{tabular}  & \begin{tabular}[c]{@{}l@{}}3.92†\\ (2.50 - 6.10)\end{tabular} & \begin{tabular}[c]{@{}l@{}}2.20†\\ (1.73 - 2.78)\end{tabular} &  & \begin{tabular}[c]{@{}l@{}}4.26†\\ (2.46 - 6.85)\end{tabular}  & \begin{tabular}[c]{@{}l@{}}3.67†\\ (2.25 - 5.82)\end{tabular}  & \begin{tabular}[c]{@{}l@{}}3.71†\\ (2.22 - 6.22)\end{tabular}  & \begin{tabular}[c]{@{}l@{}}4.17†\\ (2.45 - 6.86)\end{tabular} & \begin{tabular}[c]{@{}l@{}}4.08†\\ (2.36 - 6.39)\end{tabular}  \\
                                                       & \begin{tabular}[c]{@{}l@{}}Group 2\\ (n=25)\end{tabular}    & \begin{tabular}[c]{@{}l@{}}4.43†\\ (3.12 - 6.06)\end{tabular}  & \begin{tabular}[c]{@{}l@{}}4.74†\\ (3.31 - 6.49)\end{tabular} & \begin{tabular}[c]{@{}l@{}}2.58†\\ (1.44 - 4.38)\end{tabular} &  & \begin{tabular}[c]{@{}l@{}}3.95†\\ (2.47 - 5.62)\end{tabular}  & \begin{tabular}[c]{@{}l@{}}5.93†\\ (3.44 - 9.28)\end{tabular}  & \begin{tabular}[c]{@{}l@{}}5.02†\\ (2.67 - 8.68)\end{tabular}  & \begin{tabular}[c]{@{}l@{}}3.53†\\ (2.41 - 4.96)\end{tabular} & \begin{tabular}[c]{@{}l@{}}4.51†\\ (2.74 - 6.58)\end{tabular}  \\
                                                       & \begin{tabular}[c]{@{}l@{}}Group 3\\ (n=34)\end{tabular}    & \begin{tabular}[c]{@{}l@{}}8.93†\\ (5.60 - 12.60)\end{tabular} & \begin{tabular}[c]{@{}l@{}}9.83\\ (6.15 - 14.11)\end{tabular} & \begin{tabular}[c]{@{}l@{}}4.11†\\ (2.37 - 6.46\end{tabular}  &  & \begin{tabular}[c]{@{}l@{}}7.27†\\ (3.88 - 11.47)\end{tabular} & \begin{tabular}[c]{@{}l@{}}11.07\\ (5.44 - 18.61)\end{tabular} & \begin{tabular}[c]{@{}l@{}}7.81†\\ (4.56 - 11.65)\end{tabular} & \begin{tabular}[c]{@{}l@{}}8.89\\ (4.23 - 15.07)\end{tabular} & \begin{tabular}[c]{@{}l@{}}10.57\\ (5.18 - 17.21)\end{tabular} \\ \bottomrule
\end{tabular}}
    \label{tab:hr3_result}
    \parbox{\textwidth}{\raggedright \footnotesize  Group 1, 2, 3 correspond to Fitzpatrick skin Types I-III, IV-V, and VI, respectively.
    \\ †Met pre-specified target of MAPE $<10$\% (p $<0.05$)}
\end{table}

\newpage

\begin{table}[H]
    \centering
    \caption{\textbf{Measurement success rate by the PHRM in diverse laboratory conditions and lighting.}}
    \resizebox{\textwidth}{!}{\begin{tabular}{llllllllll}
	\toprule                                                                         & \multirow{2}{*}{Group}                                      & \multirow{2}{*}{All}                                            & \multicolumn{2}{l}{Condition}                                   & \multicolumn{5}{l}{Lighting}                                     \\
	                                                                                 &                                                             &                                                                 & At-Rest                                                         & Post-Exercise                                                   & Fluorescent                                                     & Incandescent                                                    & Natural                                                         & Dim LED                                                         & Normal LED                                                      \\
	\midrule \multirow{4}{*}{\begin{tabular}[c]{@{}l@{}}Success\\ Rate\end{tabular}} & \begin{tabular}[c]{@{}l@{}}Total\\ (n=103)\end{tabular}     & \begin{tabular}[c]{@{}l@{}}75.66\\ (69.75 - 81.22)\end{tabular} & \begin{tabular}[c]{@{}l@{}}78.39\\ (72.39 - 84.06)\end{tabular} & \begin{tabular}[c]{@{}l@{}}62.07\\ (53.52 - 70.67)\end{tabular} & \begin{tabular}[c]{@{}l@{}}85.02\\ (79.07 - 90.49)\end{tabular} & \begin{tabular}[c]{@{}l@{}}75.09\\ (68.08 - 81.70)\end{tabular} & \begin{tabular}[c]{@{}l@{}}73.96\\ (65.54 - 81.99)\end{tabular} & \begin{tabular}[c]{@{}l@{}}77.03\\ (69.09 - 84.21)\end{tabular} & \begin{tabular}[c]{@{}l@{}}80.99\\ (73.87 - 87.71)\end{tabular} \\
	                                                                                 & \begin{tabular}[c]{@{}l@{}}Skintone 1\\ (n=44)\end{tabular} & \begin{tabular}[c]{@{}l@{}}93.29\\ (89.87 - 96.06)\end{tabular} & \begin{tabular}[c]{@{}l@{}}95.88\\ (93.02 - 98.20)\end{tabular} & \begin{tabular}[c]{@{}l@{}}78.99\\ (68.60 - 88.14)\end{tabular} & \begin{tabular}[c]{@{}l@{}}97.78\\ (95.24 - 100.0)\end{tabular} & \begin{tabular}[c]{@{}l@{}}93.70\\ (88.97 - 97.66)\end{tabular} & \begin{tabular}[c]{@{}l@{}}94.57\\ (89.92 - 98.41)\end{tabular} & \begin{tabular}[c]{@{}l@{}}94.78\\ (90.30 - 98.48)\end{tabular} & \begin{tabular}[c]{@{}l@{}}98.47\\ (96.06 - 100.0)\end{tabular} \\
	                                                                                 & \begin{tabular}[c]{@{}l@{}}Skintone 2\\ (n=25)\end{tabular} & \begin{tabular}[c]{@{}l@{}}80.44\\ (70.18 - 89.22)\end{tabular} & \begin{tabular}[c]{@{}l@{}}83.33\\ (72.63 - 92.25)\end{tabular} & \begin{tabular}[c]{@{}l@{}}65.28\\ (47.83 - 80.77)\end{tabular} & \begin{tabular}[c]{@{}l@{}}87.67\\ (76.71 - 97.10)\end{tabular} & \begin{tabular}[c]{@{}l@{}}82.05\\ (70.83 - 91.67)\end{tabular} & \begin{tabular}[c]{@{}l@{}}81.25\\ (66.23 - 92.86)\end{tabular} & \begin{tabular}[c]{@{}l@{}}82.67\\ (69.41 - 94.12)\end{tabular} & \begin{tabular}[c]{@{}l@{}}83.33\\ (69.44 - 95.45)\end{tabular} \\
	                                                                                 & \begin{tabular}[c]{@{}l@{}}Skintone 3\\ (n=34)\end{tabular} & \begin{tabular}[c]{@{}l@{}}44.60\\ (36.13 - 53.35)\end{tabular} & \begin{tabular}[c]{@{}l@{}}45.85\\ (36.99 - 55.31)\end{tabular} & \begin{tabular}[c]{@{}l@{}}39.39\\ (24.24 - 54.90)\end{tabular} & \begin{tabular}[c]{@{}l@{}}60.76\\ (47.14 - 74.03)\end{tabular} & \begin{tabular}[c]{@{}l@{}}40.48\\ (27.59 - 52.88)\end{tabular} & \begin{tabular}[c]{@{}l@{}}32.91\\ (19.74 - 47.44)\end{tabular} & \begin{tabular}[c]{@{}l@{}}44.83\\ (31.11 - 59.14)\end{tabular} & \begin{tabular}[c]{@{}l@{}}50.62\\ (35.80 - 65.48)\end{tabular} \\
	\bottomrule
\end{tabular}}
    \label{tab:hr3_rate}
    \parbox{\textwidth}{\raggedright \footnotesize Group 1, 2, 3 correspond to Fitzpatrick skin Types I-III, IV-V, and VI, respectively.}
\end{table}

\clearpage
\newpage
\section{Additional Results in Free-Living Setting}

\begin{figure*}[h!]
    \vspace{1cm}
    \centering
    \includegraphics[width=0.4 \textwidth]{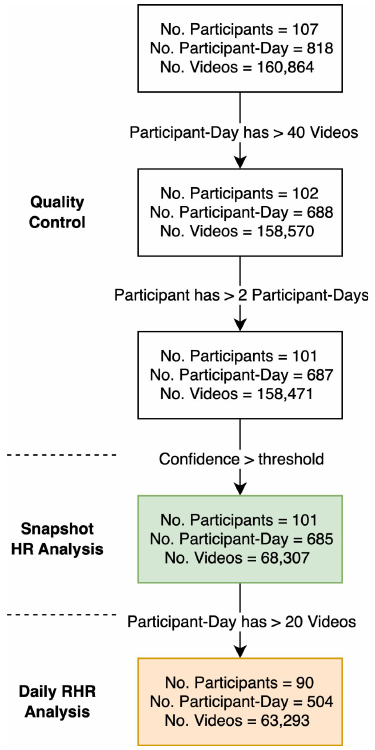}
    \caption{\textbf{Flowchart of inclusion criteria for free-living study.} We excluded 6 participants who did not meet the minimum adherence criteria of at least 3 days with more than 40 video clips per day. Valid heart rate (HR) measurements determined by gating on the deep learning model’s confidence scores associated with the HR predictions. }
    \label{fig:flow_chart} 
\end{figure*}

\begin{figure*}[h!]
    \centering
    \includegraphics[width=1 \textwidth]{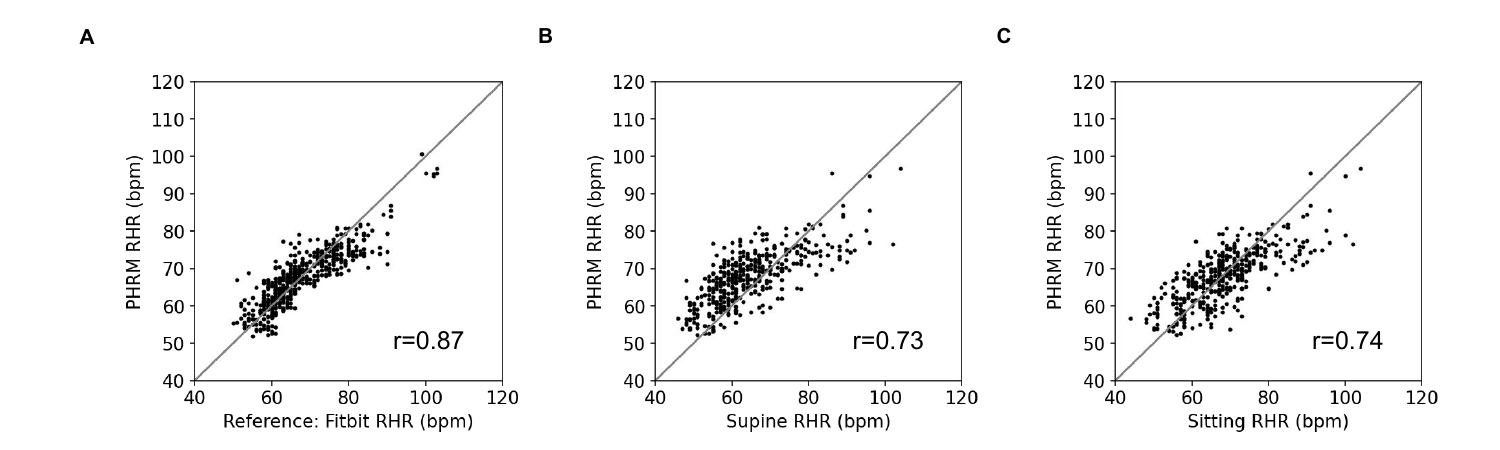}
    \caption{\textbf{Comparison of daily RHR measurements by the PHRM against reference and traditional RHR measurements in free-living conditions.} Scatter plots showing the agreement between PHRM-estimated daily RHR values and (A) reference RHR from a Fitbit wearable HR tracker, (B) supine RHR measurements from ECG, (C) sitting RHR measurements from ECG, respectively. }
    \label{fig:supp_rhr} 
\end{figure*}

\begin{figure*}[h!]
    \centering
    \includegraphics[width=1 \textwidth]{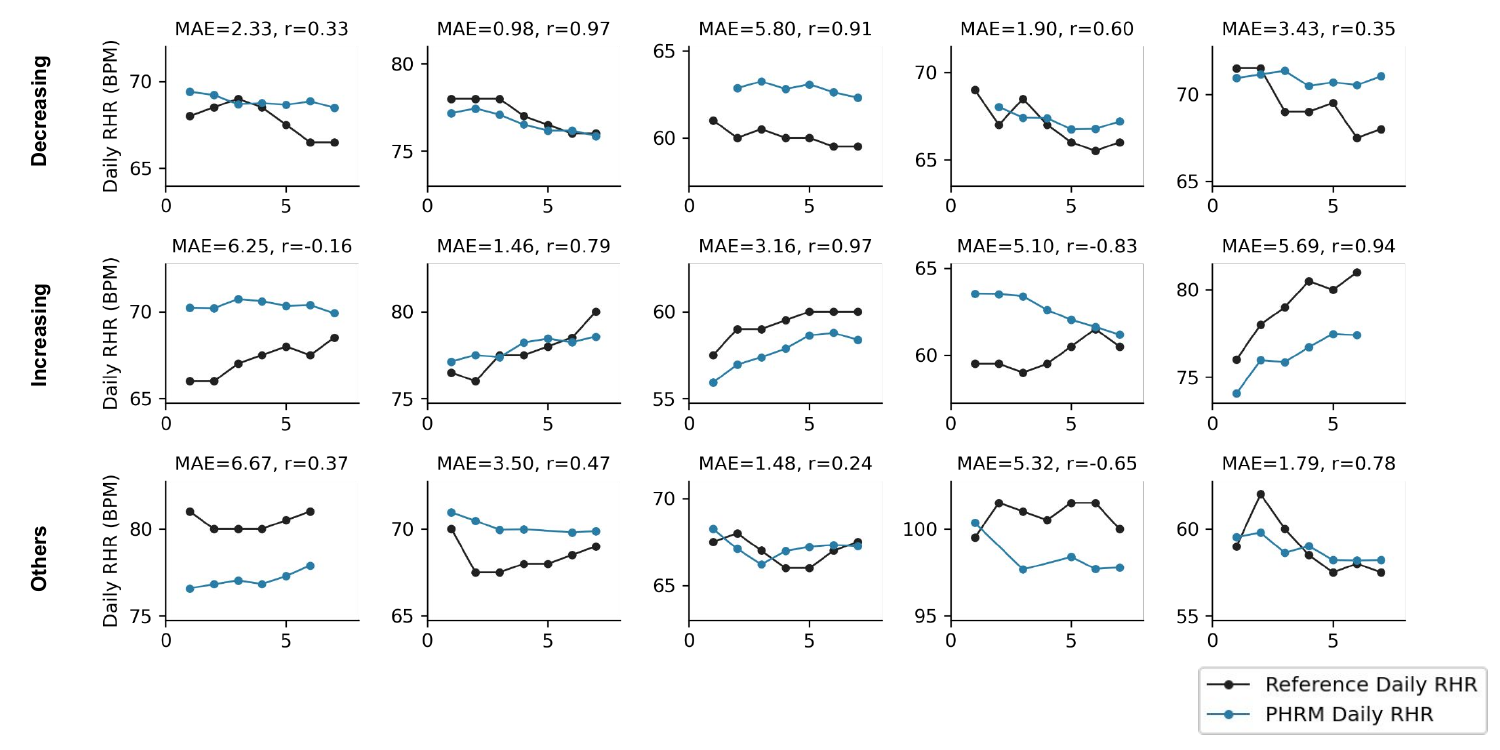}
    \caption{\textbf{Trends for daily RHR measurements over the week.} Comparison of daily RHR estimates from the proposed PHRM and a reference wearable RHR tracker over a 7-day period, sampled from participants exhibiting diverse trends. To enhance interpretability, participants were categorized into three groups based on reference wearable RHR trends: decreasing, increasing, and other patterns.  }
    \label{fig:rhr_example} 
\end{figure*}

In this section, we detail following figures that present results of our analysis in free-living conditions. Fig.~\ref{fig:flow_chart} presents the flowchart outlining the inclusion criteria applied throughout the study, including quality control measures and filtering steps for valid data. Fig.~\ref{fig:supp_rhr} compares the PHRM’s daily RHR measurements with traditional RHR values in free-living conditions, illustrating the agreement with both supine and sitting measurements. Additionally, Fig.~\ref{fig:rhr_example} highlights the trends in daily RHR measurements over a seven-day period, comparing PHRM results with those from a reference wearable heart rate tracker, offering insight into the variability of participant data.

\clearpage
\newpage
\section{PHRM Ablation Experiments}
\label{app:hr_ablation_exp}

\subsection{Video Preprocessing}
\label{app:hr_ablation_video_preprocessing}

In this section, we analyze these preprocessing techniques to address challenges related
to video stabilization, frame rate variability, resolution, and temporal
consistency, which directly impact the accuracy of HR signals. By evaluating
each technique's contribution (Table~\ref{tab:ablation_preprocessing}), we aim to
identify the optimal configuration that balances computational efficiency with signal
retention, ensuring robust performance across diverse video conditions.

\begin{table}[H]
	\centering
	\caption{\textbf{Ablation Studies of Video Preprocessing.} The baseline is the
	reported model, trained with video stabilization, a resolution of 32x32, and
	frame differencing. Gated-RMSE is the Development metric, computed as RMSE
	after excluding the bottom 20\% of videos with the lowest confidence scores.
	For each row, a separate hyper-parameter search was performed, and the best
	tune metric is presented.}
	\resizebox{0.35 \textwidth}{!}{\begin{tabular}{lc}
\toprule
                       & Gated RMSE              \\
\midrule
Baseline    & 10.37                   \\
48x48 Resolution       & 10.52                   \\
16x16 Resolution       & NaN \\
No Video Stabilization & 10.77                   \\
No Frame Differencing  & 12.08                  \\ \bottomrule
\end{tabular}} \label{tab:ablation_preprocessing}
\end{table}

\begin{figure*}[h!]
	\centering
	\includegraphics[width=0.45 \textwidth]{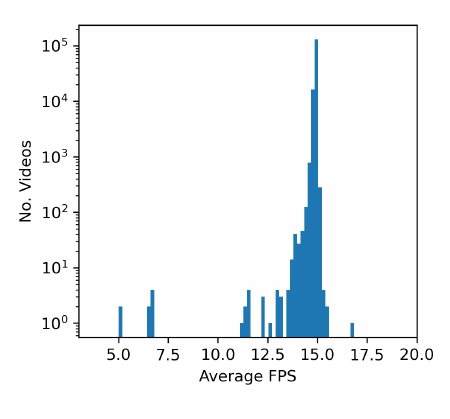}
	\caption{\textbf{Video FPS on Freeliving Test Cohort}.This figure shows the distribution
	of average FPS across videos, with the majority of videos falling within the
	12 to 15 FPS range.}
	\label{fig:freeliving_fps}
\end{figure*}

\begin{figure*}[h!]
	\centering
	\includegraphics[width=1 \textwidth]{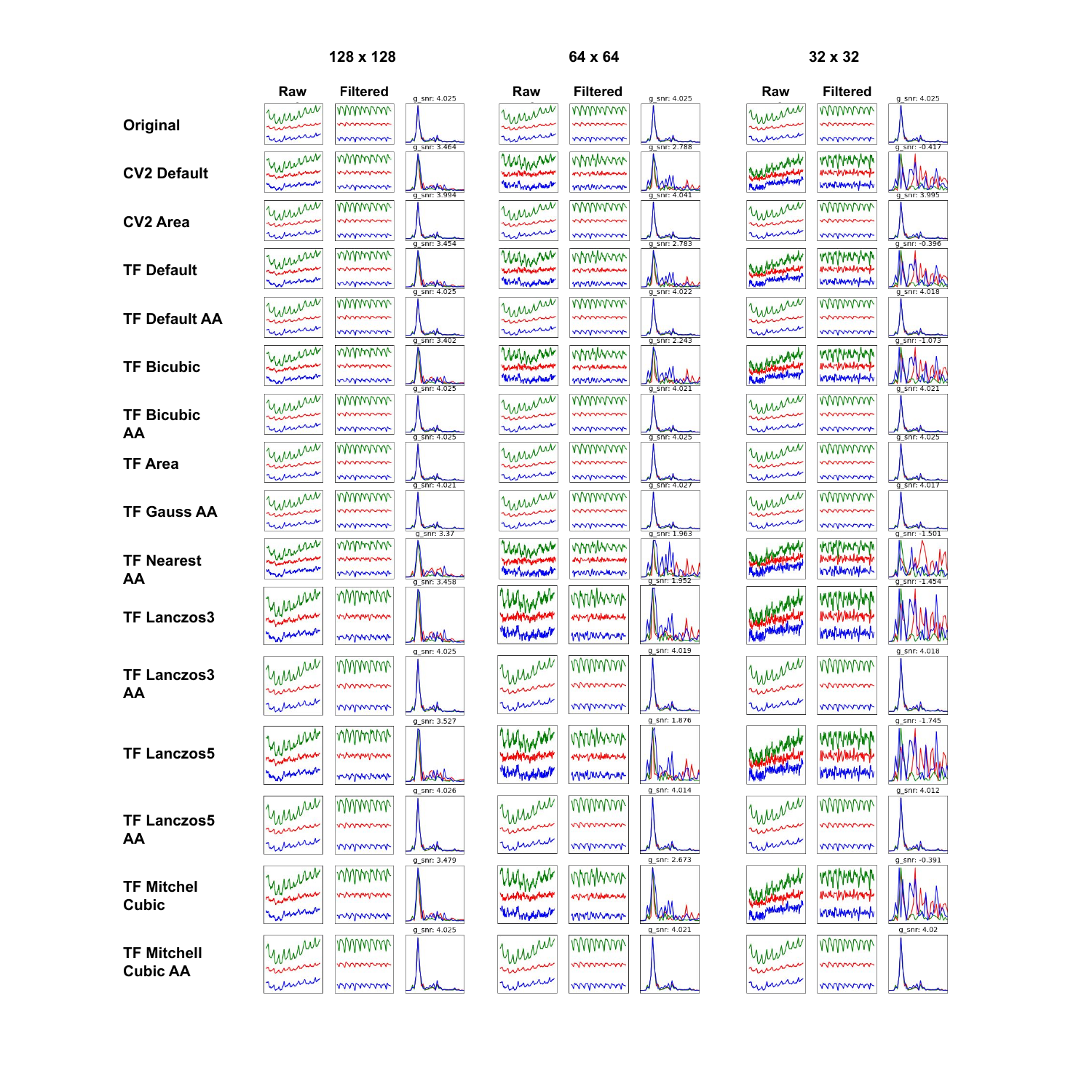}
	\caption{\textbf{Different Resizing Methods and Resolutions.} Comparison of raw
	and processed signals using different resizing methods and resolutions (128x128,
	64x64, 32x32). Each row represents a different resizing method, including OpenCV
	(CV2) and TensorFlow (TF) variants, with and without anti-aliasing (AA). The columns
	display raw signals, detrended and filtered signals, and the corresponding signal-to-noise
	ratio (SNR) plots. The SNR values highlight the impact of resolution and
	resizing techniques on signal quality, with higher resolutions generally maintaining
	stronger signal integrity across different methods.}
	\label{fig:resize_method}
\end{figure*}

\begin{itemize}
	\item \textbf{Video Stabilization.} In Table~\ref{tab:ablation_preprocessing},
		we compare the performance of models with and without video stabilization on
		the tune set. Video stabilization improved gated-RMSE by 0.40. However,
		while it enhanced overall performance, we observed that stabilization failed
		in some cases and introduced noticeable artifacts in certain videos. This
		presents an opportunity for further improvements.

	\item \textbf{Resizing.} Video resolution is critical for balancing
		computational cost and HR signal retention. Our goal was to identify the smallest
		video resolution that retains HR information and apply it consistently across
		frames. Fig.~\ref{fig:resize_method} compares 15 resizing methods from OpenCV2
		and TensorFlow across three resolutions: 128x128, 64x64, and 32x32. In the illustration,
		the original video has a green channel SNR of 4.025. At higher resolutions (e.g.,
		128x128), the resizing method has little effect on HR signal retention, with
		SNR values ranging from 3.37 to 4.025. However, as resolution decreases, the
		choice of resizing method becomes more significant. At 32x32, TensorFlow’s
		area resizing method maintained an SNR of 4.025, matching the original video,
		making it the preferred method. Additionally, Table~\ref{tab:ablation_preprocessing}
		compares the impact of different resolutions (48x48 and 16x16) on model
		training. We found that 32x32 was the minimal resolution that did not degrade
		performance on the tune set.

	\item \textbf{Frame Differencing.} Table~\ref{tab:ablation_preprocessing}
		shows that removing frame differencing reduces performance on the tune set
		from 9.55 to 12.08.

	\item \textbf{Linear Interpolation}. Fig.~\ref{fig:freeliving_fps} shows the
		average video frames-per-second (FPS) distribution for the free-living test
		dataset, with an average FPS of 14.94 and a standard deviation of 0.10.
		Notably, over 76.4\% of videos have an FPS lower than 15, underscoring the need
		to standardize the frame rate.
\end{itemize}

\clearpage
\newpage

\subsection{HR Network}
\label{app:hr_ablation_network}

We conducted comprehensive ablation experiments to evaluate the effectiveness of
each component in the deep model. In each experiment, a single component was
modified to assess its impact on performance. The full list of ablation experiments
is provided below, with performance on the tune set for each variation shown in
Table~\ref{tab:ablation_network}.

\begin{table}[H]
	\centering
	\caption{\textbf{Ablation Studies of rPPG HR Network.} The table presents results
	on the tune set, showing how modifications to the network backbone, HR output head,
	loss functions, data augmentation techniques, and ensemble strategies affect
	the model's accuracy. Each ablation study isolates a single component to assess
	its contribution to overall performance.}
	\resizebox{0.55 \textwidth}{!}{\begin{tabular}{llc}
\toprule
\multicolumn{2}{l}{}                                              & Gated RMSE \\ \midrule
\multicolumn{2}{l}{Baseline (No Ensemble)}                        & 10.37      \\
\multirow{2}{*}{Architecture}         & Smaller Model             & 10.91      \\
                                      & Larger Model              & 10.42      \\
\multirow{3}{*}{Output Head and Loss} & FFT Head and MAPE Loss    & 11.02      \\
                                      & Dense Head and MAPE Loss  & 11.63      \\
                                      & Dense Head and Focal Loss & 12.75      \\
\multirow{3}{*}{Augmentations}        & No Rotation               & 10.69      \\
                                      & No Cropping and Resizing  & 10.61      \\
                                      & No Speed Augmentation     & 10.79      \\
\multirow{2}{*}{Ensemble}             & Top 3                     & 9.96       \\
                                      & Top 5                     & 9.92      \\ \bottomrule
\end{tabular}} \label{tab:ablation_network}
\end{table}

\begin{itemize}
	\item \textbf{Network Backbone.} We tested both smaller and larger model
		configurations. Our model outperformed the smaller configuration and matched
		the performance of the larger one.
		\begin{itemize}
			\item Smaller model: Channel sizes of 2, 8, 16, 32 and 64.

			\item Larger model: Channel sizes of 8, 32, 64, 128 and 258.
		\end{itemize}

	\item \textbf{HR Output Head and Loss Function.} We found the model to be
		sensitive to the choice of HR output and loss function. As described in the Methods
		section, the reported approach (FFT head + Focal loss) treats HR extraction
		as a classification task. We also tested different combinations of output head
		and loss function:
		\begin{itemize}
			\item Dense Head + Focal Loss: Replaces the FFT with a traditional dense
				layer and softmax to map the pseudo-PPG to HR bins, using focal loss.

			\item Dense Head + MAPE Loss: Uses a dense layer to directly output a
				single HR value from the pseudo-PPG.

			\item FFT Head + MAPE Loss: Applies a weighted sum
				of HR probabilities and HR bins, using MAPE loss between predicted and ground
				truth HR values.
		\end{itemize}

	\item \textbf{Data Augmentation.} We evaluated the impact of removing each
		augmentation method on tune performance. Each contributed between 0.30 and 0.42
		gated-RMSE points.

		\begin{itemize}
			\item No Rotation: Removes rotation.

			\item No Cropping and Resizing: Removes cropping and resizing.

			\item No Speed Augmentation: Removes speed augmentation.
		\end{itemize}

	\item \textbf{Ensemble.} Lastly, we compared various ensemble strategies:
		\begin{itemize}
			\item Top 3 Checkpoints: Ensembles predictions from the top 3 models.

			\item Top 5 Checkpoints: Ensembles predictions from the top 5 models.
		\end{itemize}
\end{itemize}
\clearpage
\newpage

\subsection{Confidence Gating}
\label{app:hr_ablation_video_gating}

We evaluated two alternatives for obtaining confidence scores: the maximum of HR
probabilities and the pseudo-PPG SNR. Table~\ref{tab:ablation_gating} compares
their performance on the tune dataset. When the acceptance rate was set to 80\%,
negative entropy demonstrated superior performance over both alternatives.

\begin{table}[H]
	\centering
	\caption{\textbf{Confidence Gating Alternatives.} This table presents the performance
	of two confidence scoring methods—maximum of HR probabilities and pseudo-PPG SNR—on
	the tune dataset. Negative entropy demonstrated superior results when the
	acceptance rate was set to 80\%.}
	\resizebox{0.5 \textwidth}{!}{\begin{tabular}{llc}
\toprule
\multicolumn{2}{l}{}                                        & Gated RMSE \\ \midrule
Baseline     &            & 9.92       \\
\multirow{2}{*}{Alternatives} & Maximum of HR Probabilities & 10.04      \\
                              & Pseudo PPG SNR              & 11.53     \\ \bottomrule
\end{tabular}} \label{tab:ablation_gating}
\end{table}

\begin{figure*}[h!]
	\centering
	\includegraphics[width=0.9 \textwidth]{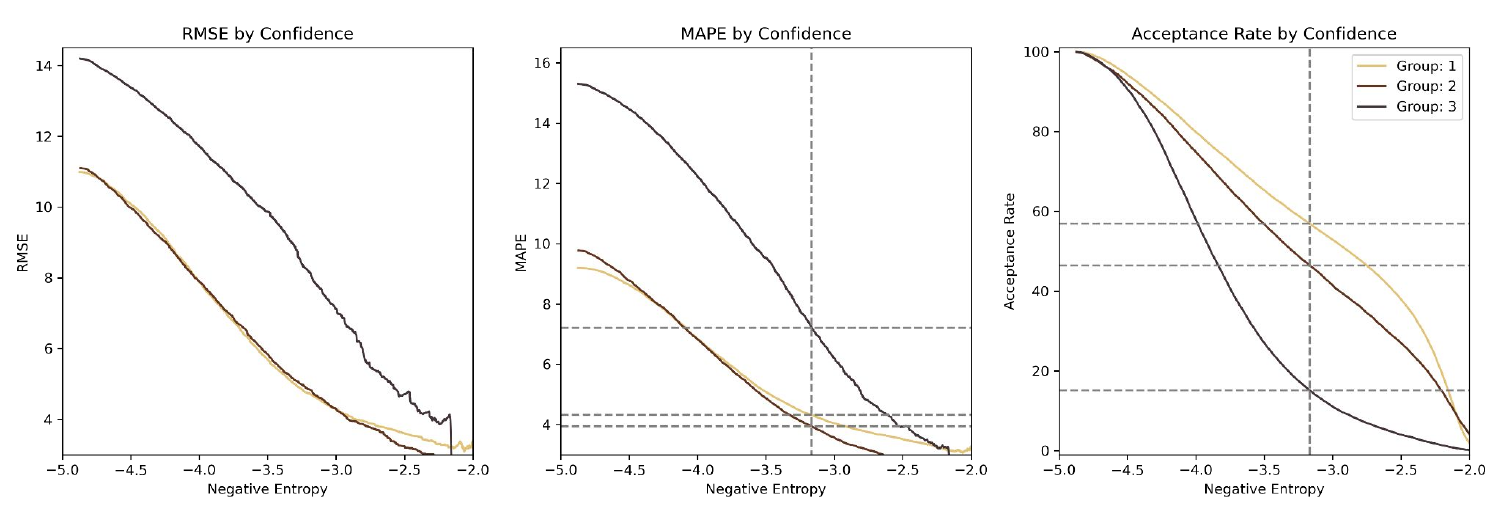}
	\caption{\textbf{Confidence Gating Threshold Search.} This figure illustrates the
	process of determining the optimal confidence threshold using the free-living tune
	dataset. The negative entropy threshold was set at -3.17 (indicated by the
	gray line) to ensure that the MAPE for each skin tone group remained below 8 and
	the MAPE difference between groups was less than 3.}
	\label{fig:gating_search}
\end{figure*}

Fig.~\ref{fig:gating_search} illustrates the process of determining the confidence
threshold using the free-living tune dataset. As outlined in the Methods section,
the goal was to set a threshold where the MAPE for each skin tone group remains below
8, and the MAPE difference between groups is less than 3. Based on this
criterion, we selected a negative entropy threshold of -3.17, highlighted by the
gray line in the figure.

\clearpage
\newpage

\subsection{Hyper-Parameters}
\label{app:hr_ablation_hyperparameter}

This subsection details the hyper-parameter choices that were chosen by search algorithm. 

\begin{table}[H]
	\centering
	\caption{\textbf{Hyper-parameter Choices.} We presents the hyper-parameters for top 5 runs in Tune dataset, and these runs were chosen automatically by searching algorithm.} 
	\resizebox{0.9 \textwidth}{!}{\begin{tabular}{llllllll}
\toprule
\multicolumn{2}{c}{random\_crop\_with\_resize} & random\_flip & \multicolumn{3}{c}{random\_speed}    & \multicolumn{2}{c}{Optimizer} \\
probablity         & minimal crop size         & probablity   & min\_speed & max\_speed & probablity & batch size   & learning rate   \\ \midrule
7.63109            & 0.577256                  & 0.412314     & 0.85       & 1.15       & 1          & 32           & 0.003           \\
7.64416            & 0.615019                  & 0.406043     & 0.85       & 1.154268   & 1          & 32           & 0.003           \\
7.624905           & 0.752059                  & 0.300096     & 0.85       & 1.15       & 1          & 32           & 0.003           \\
7.697844           & 0.557755                  & 0.445152     & 0.85       & 1.15       & 1          & 64           & 0.003           \\
7.655166           & 0.580071                  & 0.408653     & 0.849836   & 1.15       & 0.999916   & 32           & 0.003          \\ \bottomrule
\end{tabular}} \label{tab:hyper-parameter}

\end{table}

\end{document}